# Geologic Constraints on Early Mars Climate

Edwin S. Kite, University of Chicago (kite@uchicago.edu).

Accepted by *Space Science Reviews.*

**Abstract.** Early Mars climate research has well-defined goals (Mars Exploration Program Analysis Group 2018). Achieving these goals requires geologists and climate modelers to coordinate. Coordination is easier if results are expressed in terms of well-defined parameters. Key parameters include the following quantitative geologic constraints. (1) Cumulative post-3.4 Ga precipitation-sourced water runoff in some places exceeded 1 km column. (2) There is no single Early Mars climate problem: the traces of ≥2 river-forming periods are seen. Relative to rivers that formed earlier in Mars history, rivers that formed later in Mars history are found preferentially at lower elevations, and show a stronger dependence on latitude. (3) The duration of the longest individual river-forming climate was >($10^2$ - $10^3$) yr, based on paleolake hydrology. (4) Peak runoff production was >0.1 mm/hr. However, (5) peak runoff production was intermittent, sustained (in a given catchment) for only <10% of the duration of river-forming climates. (6) The cumulative number of wet years during the valley-network-forming period was >$10^5$ yr. (7) Post-Noachian light-toned, layered sedimentary rocks took >$10^7$ yr to accumulate. However, (8) an "average" place on Mars saw water for <$10^7$ yr after the Noachian, suggesting that the river-forming climates were interspersed with long globally-dry intervals. (9) Geologic proxies for Early Mars atmospheric pressure indicate pressure was not less than 0.012 bar but not much more than 1 bar. A truth table of these geologic constraints versus currently published climate models shows that the late persistence of river-forming climates, combined with the long duration of individual lake-forming climates, is a challenge for most models.

## 1. Introduction

Explaining rivers and lakes on Early Mars is difficult. The chief difficulty is that Mars 3.5 Ga received just ⅓ of the modern Earth's insolation (Haberle et al. 1998, Bahcall et al. 2001). Such low insolation puts Early Mars outside the circumstellar habitable zone – at least according to basic forward models of habitable-planet climate (Kasting et al. 1993, Kopparapu et al. 2013). Those basic models combine the greenhouse effect of $CO_2$ and $H_2O_{(v)}$. But maximum $CO_2$ + $H_2O_{(v)}$ warming is too cold for $H_2O_{(l)}$ rivers (Forget et al. 2013, Wordsworth et al. 2013, Turbet & Tran 2017). Thus, data for Early Mars – the only geologic record that can give an independent test of Earth-derived models of planetary habitability (Ehlmann et al. 2016) – shows that those models do not work (Grotzinger et al. 2014, Dietrich et al. 2017, Vasavada 2017, Haberle et al. 2017). Because basic models do not work, recent explanations for rivers and lakes on Early Mars span a wide range of trigger mechanisms, timescales, and temperatures (e.g. Wordsworth 2016, Kite et al. 2013a, Urata & Toon 2013, Ramirez et al. 2014, Halevy & Head 2014, Kerber et al. 2015, Batalha et al. 2016, Wordsworth et al. 2017, Kite et al. 2017a, Haberle et al. 2017, Palumbo & Head 2018, Tosca et al. 2018). This is an embarrassment of riches. To test these ideas, we need more constraints than just the existence of rivers.

Fortunately, recent work gives better constraints on the climates that caused river flow on early Mars. The new work constrains the number, spatial patchiness, hydrology, and timescales of past wet climates. The new geologic target is no longer "how could Mars be warm enough for lakes?" (Squyres 1989, Carr 2007) but instead "what mechanisms best explain the trends, rhythms and aberrations in Early Mars climate that are recorded by Mars' geology?" There are also new estimates of paleo-atmospheric pressure, and steps towards paleo-temperature constraints. All these constraints are summed up in Table 1.

In this paper, I focus on quantitative summary parameters that can be used as input or test data for numerical models, both of Mars climate and also of Mars climate evolution (e.g. Armstrong et al. 2004, Wordsworth et al. 2015, von Paris et al. 2015, Manning et al. 2006, Kurahashi-Nakamura & Tajika 2006, Hu et al. 2015, Pham & Karatekin 2016, Mansfield et al. 2018). For each constraint, I state a subjective confidence level. Despite progress, today's data are still not enough to pick out one best-fit model (Table 2). Perhaps the true explanation has not yet been dreamt of.

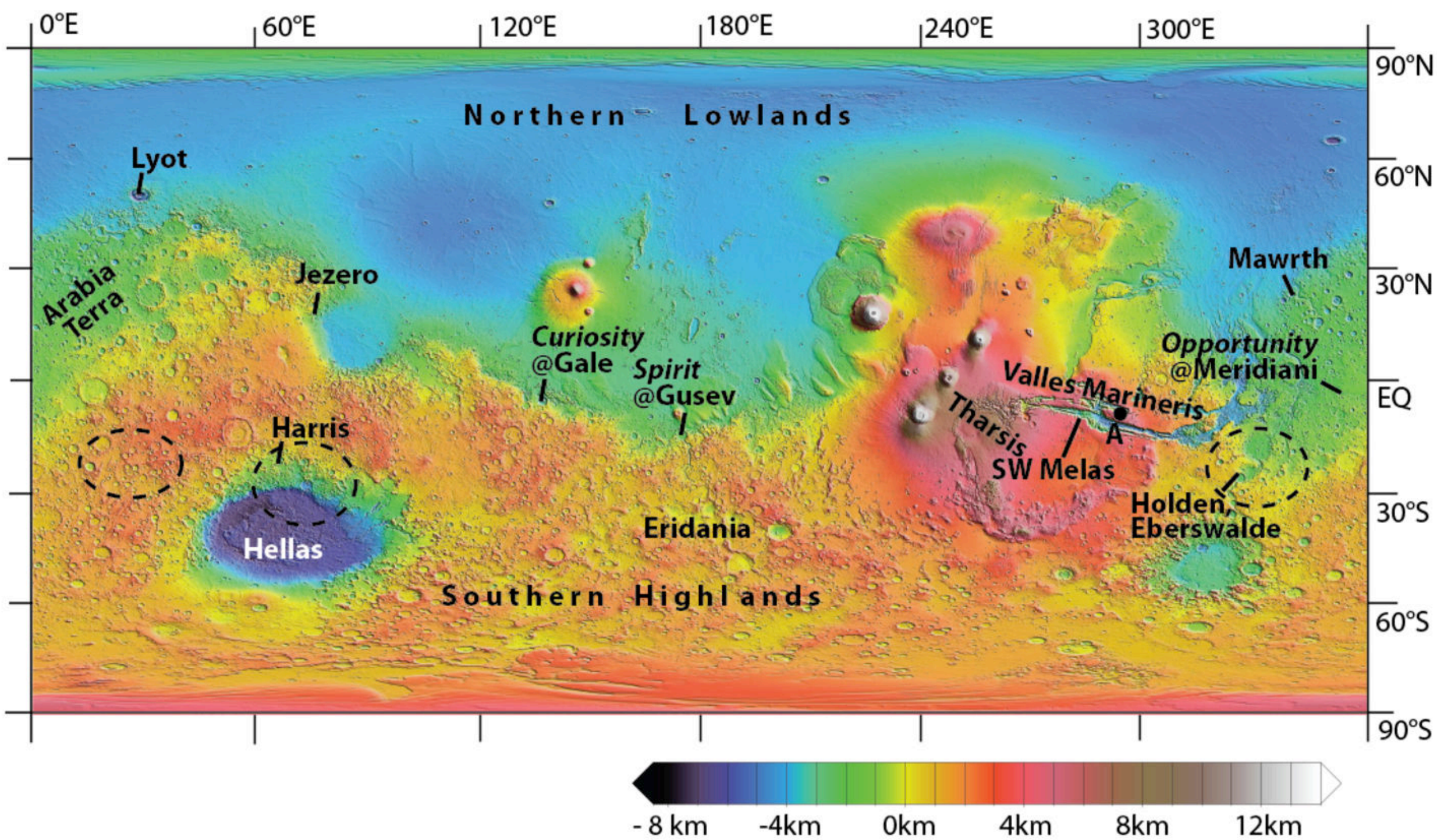

**Fig. 1.** Map of Mars showing places named in this paper. Dashed ellipses contain most of the large alluvial fans (Kraal et al. 2008a). A = thickest known sedimentary rock accumulation on Mars (easternmost mound within East Candor Chasma, 8 km thick, 8°S 66°W). Background is Mars Orbiter Laser Altimeter (MOLA; Smith et al. 2001) topography draped over shaded relief. 1° latitude ≈ 59 km.

| Parameter | Confidence level* | Constraint | Method / notes |
|---|---|---|---|
| *River-forming periods* | | | |
| Post-3.4 Ga precipitation-sourced water column (rain and/or snowmelt) | High | >1 km | §2.1, Fig. 2. Eroded-sediment thickness × minimum water/sediment ratio. |
| Time spanned by river-forming climates | High | $\gg 10^8$ yr | §2.2, Fig. 3. Exhumed-crater frequency, and differences in crater retention age. |
| Number of river-forming periods | High | ≥ 2 | §2.2, Fig. 3. Crosscutting relationships; crater counts. |
| Ocean size at maximum extent | Very high | $>1.1 \times 10^6$ km$^2$ | §2.9. Eridania sea; geomorphology. |
| Last year of global climate permitting river flow | High | <3.4 Ga | §2.2. Uncommon supraglacial channels formed <1 Ga (e.g. Dickson et al. 2009) |
| Trends between wet periods (§2.3) | Medium/ high | §2.3, Figs. 4-5. | |
| *τ-R-I-N parameters* | | | |
| Duration of individual river-forming climates ($\tau$) | Medium | > ($10^2$ – $10^3$) yr for longest climate | §2.5, Fig. 7. Lake hydrology. |
| | Low/ medium | > $2 \times 10^4$ yr for longest climate | §2.5. mm-scale laminations, interpreted as annual varves |
| Peak runoff production ($R$) | High | >0.1 mm/hr | §2.6. River discharge (inferred from paleo-channel dimensions), divided by catchment drainage area |
| Intermittency during wet events ($I$) | High | Peak runoff production <10% of the time | §2.7, Fig. 8. To avoid over-topping closed basin lakes. |
| Cumulative wet years during valley-network-forming climate episode ($\tau \times N$) | Medium | $>10^5$ yr | §2.8. Sediment transport calculations constrained by paleochannel dimensions |
| *Sedimentary-rock-forming climates* | | | |
| Duration of post-Noachian surface liquid water for "average" Mars | Medium/ high | $<10^7$ yr | §3.3, Fig 7. Persistence of easily-dissolved minerals, such as olivine |
| Years of sediment deposition in sedimentary rock record | High | $>10^7$ yr | §3.1. Counts of orbitally-paced rhythmite layers. |
| Time span of deposition for layered, indurated, equatorial sediments | High | $\gg 10^8$ yr | §3.1. Sediment deposition need not have been continuous. |
| Water column required to indurate sedimentary rocks | Medium/ high | >20 km | §3.2. Geochemical reaction-transport models. >1 km column-$H_2O$-equivalent H content in sedimentary rocks today. |
| *Pressure and temperature* | | | |
| Paleo-atmospheric pressure | Low/ medium | >0.01 bar, <(1-2) bar | §4.1. High uncertainty. |
| Peak annual-mean warm-climate temperature at river locations ($T_{av}$) | Low | -18°C to 40°C | §4.2. Poorly constrained. |

**Table 1.** Summary of key parameters for Early Mars climate research. *Confidence levels are subjective.

## 2. River-forming climates

Sediment transport by precipitation-sourced water runoff marks Mars history's wettest climates. These wettest climates are the most challenging parts of Mars climate history to explain.

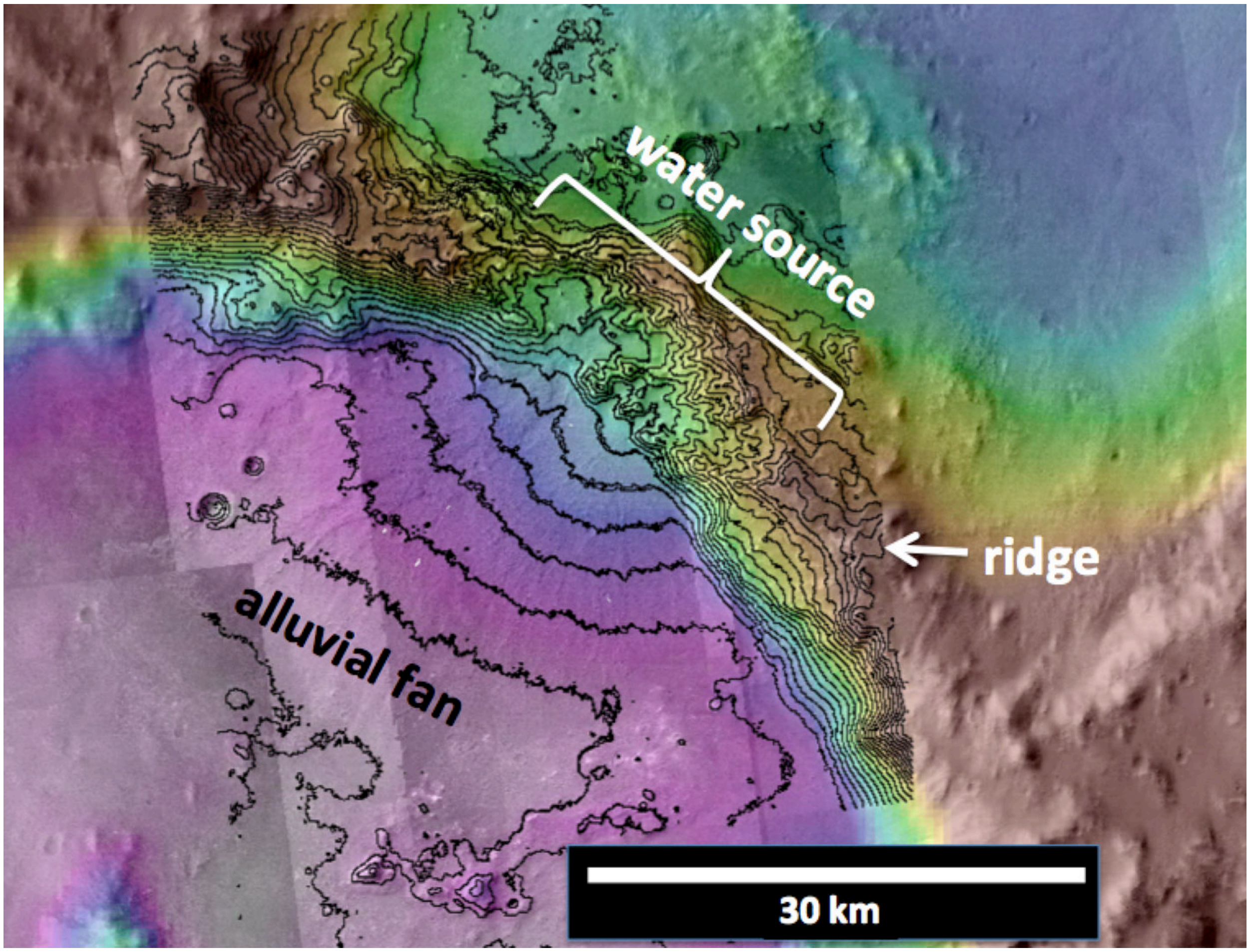

**Fig. 2.** NE rim of Harris crater (21.5°S 67°E), showing an Early Mars alluvial fan that was fed by precipitation-sourced runoff (Williams et al. 2011). 100m contour spacing on Digital Terrain Models (DTM) generated by David P. Mayer using the Ames Stereo Pipeline (Moratto et al. 2010) and images from Context Camera (CTX; Malin et al. 2007). Background is MOLA topography overlain on THEMIS (Thermal Emission Imaging System; Christensen et al. 2004) visible-light mosaic (Edwards et al. 2011).

### 2.1. Precipitation-sourced water runoff: > 1 km column post - 3.4 Ga (high confidence)

A lot of the runoff on Early Mars came from rain/snowmelt (precipitation). Observations at Harris crater (Fig. 2), and at many other sites on Mars, show channel heads <1 km from a ridge (Hynek & Phillips 2003, Williams et al. 2011, Malin et al. 2010). At 24.1°S 28.2°E, 22.7°S 73.8°E, and 19.9°S 32.7°W, channel heads are located close to a ridgeline but on opposite sides of a ridge. These data are inconsistent with spring discharge, but consistent with precipitation-sourced runoff (Mangold et al. 2004, 2008; Weitz et al. 2010). The existence of rain/snowmelt is a constraint on past climate. A wetter climate, with $T > 0$ °C at least seasonally, is required to explain these observations. A rain- or snowmelt-permitting climate almost certainly requires atmospheric pressure higher than the ~6 mbar atmospheric pressure on today's Mars, because evaporitic cooling efficiently suppresses runoff at 6 mbar (Hecht 2002, Mansfield et al. 2018). Therefore, this paper will focus on these precipitation-sourced runoff sites.

How much rain/snowmelt is required? Cumulative rain/snowmelt is estimated by first dividing fluvial sediment-deposit volume by the area of the sediment source region. That gives eroded-sediment thickness. Eroded-sediment thicknesses of ~1 km are widespread, with most km-thick alluvial fan deposits being <3.4 Ga in age (Grant & Wilson 2011, Morgan et al. 2014, Kite et al. 2017). This result is multiplied by water:sediment ratio (Williams et al. 2011). Fan morphology indicates water:sediment volume ratio >1:1. The result is a conservative lower bound on post-3.4 Ga column runoff production: >1 km (Williams et al. 2011, Dietrich et al. 2017).

This paper will focus on precipitation-sourced fluvial activity. However, runoff from precipitation was not the only cause of fluvial sediment transport on Mars. Indeed, impact-initiated volatile release cut valleys and moved sediment on Mars (e.g. El-Maary et al. 2013, Grant & Wilson 2018). Some canyons were cut by discharge from a subsurface water source (e.g. Gulick 2001, Leask et al. 2007, Burr et al. 2009). Some stubby-headed canyons have a disputed origin, with arguments in favor of formation by surface runoff (e.g. Lamb et al. 2008), and arguments in favor of a subsurface water source (e.g. Kraal et al. 2008b). Discharge of water from the subsurface is much easier when surface temperature is > 0 °C, because frozen ground traps water. However, this preference for > 0 °C is not absolute. To the contrary, groundwater discharge may continue – thanks to salinity and heat advection – even when the surface is cold and otherwise dry (Andersen 2002, Grasby et al. 2014, Scheidegger & Bense 2014, Ward & Pollard 2018, Mellon & Phillips 2001). Therefore, groundwater discharge does not, by itself, place strong constraints on climate.

### 2.2. There is no single Early Mars climate problem: rivers formed during ≥2 distinct periods (high confidence).

A river-forming climate is defined in this paper as a >10 yr interval during which precipitation-sourced runoff occurred during most years. With this definition, on Early Mars there were multiple river-forming climates (Fig. 3).

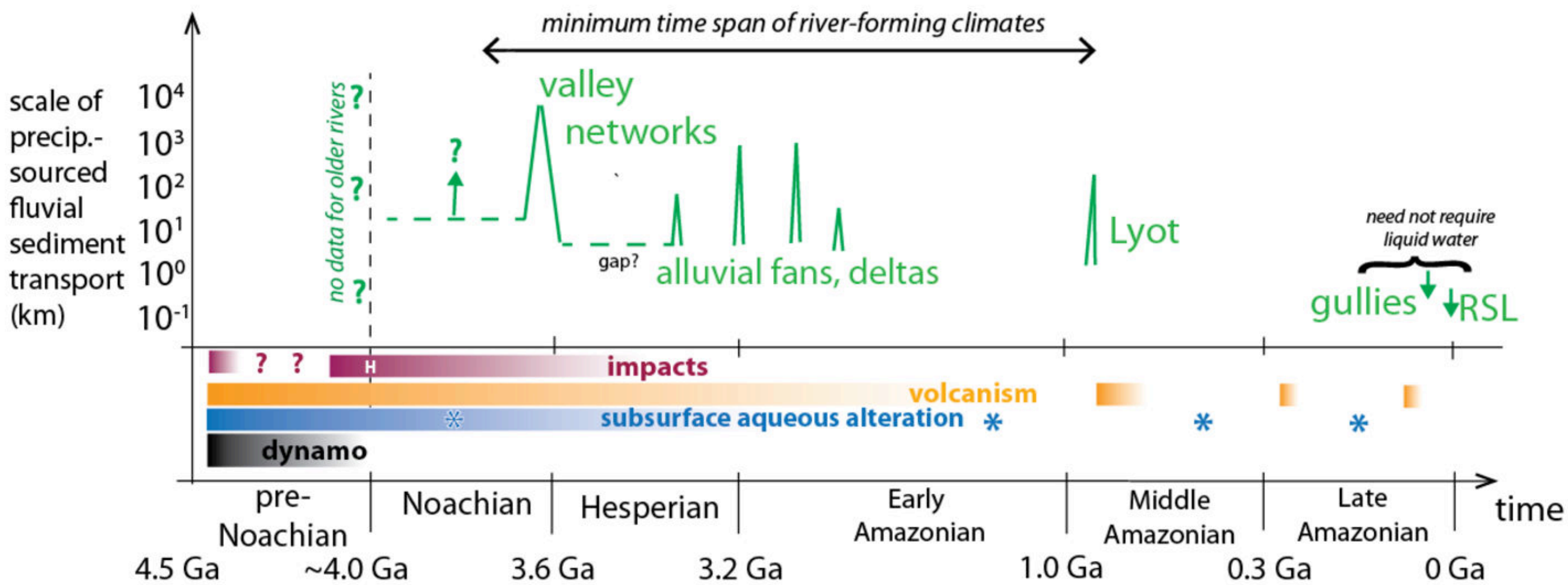

**Fig. 3.** History of Mars' river-forming climates (modified after Kite et al. 2017b). Y-axis corresponds to the map-view scale of the landforms shown. Neither the durations of geologic eras, nor the durations of river-forming climates, are to scale. Data are consistent with long globally-dry intervals. Dynamo timing is from Lillis et al. (2013). H = Hellas impact event. * = subsurface aqueous alteration as recorded by Mars meteorites (Borg & Drake 2005, Nemachin et al. 2014).

***Early/Middle Noachian (3.9 Ga – 3.7 Ga?)*[1]*:*** The first occurred after the Hellas impact[2], but before the valley networks. In this time window, craters were modified by erosion and deposition. Craters could not have been modified in one short pulse, in part because the most degraded craters tend to be older than slightly less degraded craters (Craddock & Howard 2002, Forsberg-Taylor et al. 2004, Quantin-Nataf et al. 2019). The erosion and deposition must be at least partly due to rivers and streams, and may be mostly due to rivers and streams (Craddock & Maxwell 1993, Forsberg-Taylor et al. 2004). The simplest explanation is a global climate that permitted snowmelt or even rain (Craddock & Howard 2002). This first river-forming period (Early/Middle Noachian) is clearly distinct in the geologic record from the later period of valley network formation around the Noachian/Hesperian boundary (Irwin et al. 2005a, Howard et al. 2005; Fig. 4).

***Late Noachian / Early Hesperian (~3.6 Ga):*** Across the Mars highlands around the Noachian / Hesperian boundary, regionally-integrated valley networks formed (Fassett & Head 2008a, Hynek et al. 2010, Fassett & Head 2011). These valley networks are the most obvious evidence for a warmer, wetter past on Mars (Masursky 1973). Valleys connect paleolakes over water flow paths >$10^3$ km long, which thread most of the Southern

---

[1] Absolute date estimates in this paper are given in Appendix A. That chronology is based on radiometrically-dated Lunar samples, extrapolated using crater counts to Mars. But these ages have big error bars (e.g. Robbins, 2014). In-situ radiometric ages for Mars samples are now being acquired using the *Curiosity* rover (Farley et al 2014). However, so far, these ages have not bfeen securely correlated to the crater-density age of any terrain.

[2] We take the start point for Mars' legible-from-orbit record of climate change to be the Hellas impact (Smith et al. 1999). Pre-Hellas climate history may be found in megabreccia, rare chunks of uplifted ancient crust, and possibly in meteorites (Humayan et al. 2013, Cannon et al. 2017).

Highlands (Irwin et al. 2005a, Howard et al. 2005, Fassett & Head 2008b, Goudge et al. 2012). This regional integration of Mars watersheds, which involved >$10^5$ yr of wet climate (§2.8), was never repeated.

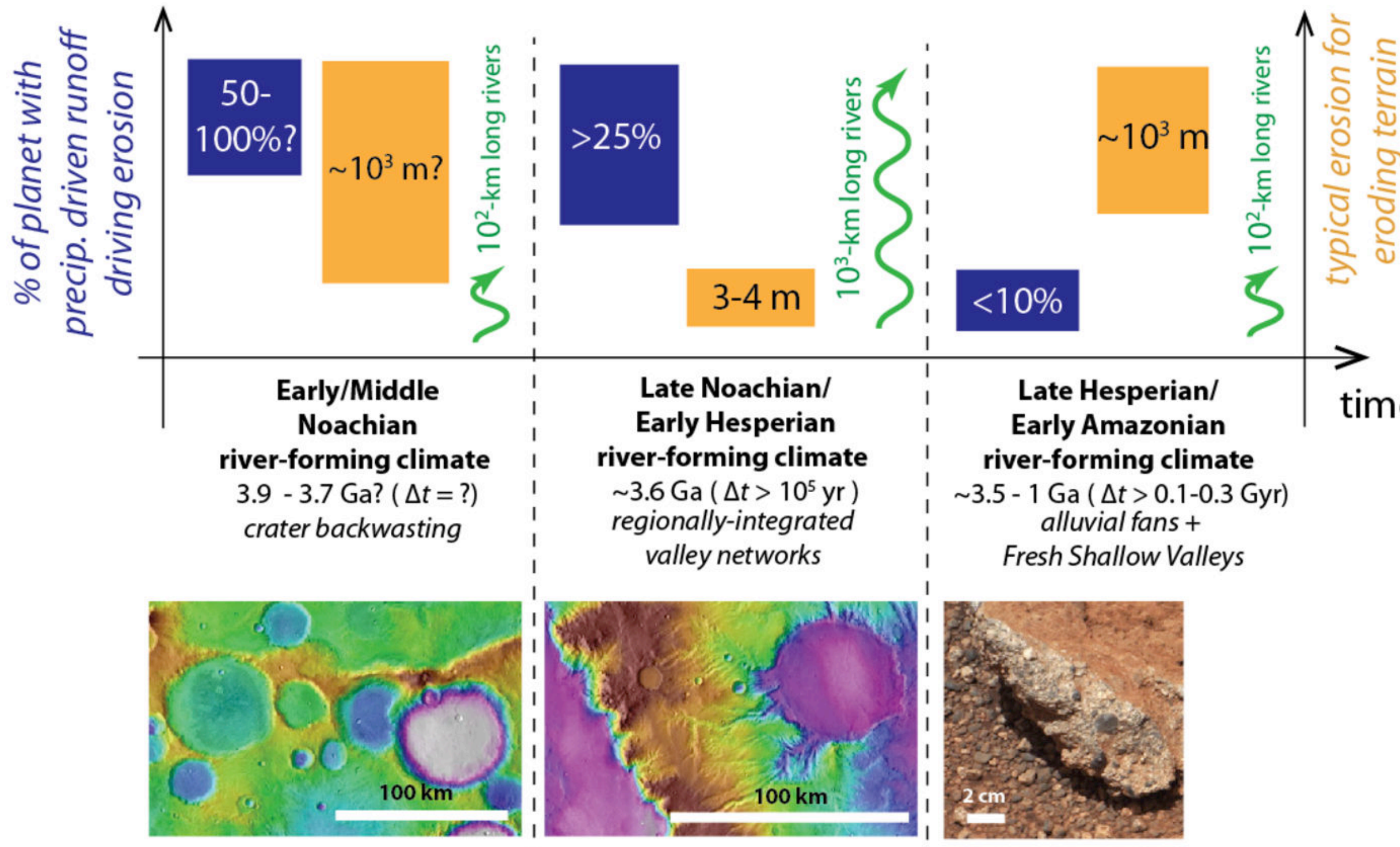

**Fig. 4.** Distinct river-forming periods show different spatial coverage, different river lengths, and different amounts of erosion by fluvial sediment transport. **Early/Middle Noachian image:** Crater backwasting. Centered near 43.5°E, 20.3°S. Image is 185 km across and color scale runs from +0.8km to +3.5 km. (same region as in Craddock & Howard 2002, their Fig. 8). **Late Noachian/Early Hesperian image:** Image is 137 km across and color scale runs from +1km to +2.5 km (for discussion see Howard 2007, his Fig. 14). Centered near 60°E, 12.5°S. **Late Hesperian / Amazonian image:** Fluvial conglomerate at Gale crater, seen close-up by *Curiosity* (Williams et al. 2013).

***Late Hesperian / Amazonian (~3.4 Ga - ~1 Ga):*** Instead, following an interval of deep wind erosion indicating dry conditions (Zabrusky et al. 2012), the Late Hesperian and Amazonian saw runoff forming closed-basin lakes and alluvial fans (Moore & Howard 2005; Kraal et al. 2008a; Grant & Wilson 2011, 2012; Morgan et al. 2014; Palucis et al. 2016; Mangold et al. 2012; Milliken et al. 2014; Goudge et al. 2016; Kite et al. 2015). Fans formed over a time span of >(100-300) Myr (Kite et al. 2017b), but fan sediment deposit thickness locally reach ~1km. Dividing ~1km sediment thickness by time span >(100-300) Myr gives average fan accumulation rate $\lesssim$10 μm/yr. Such a slow rate suggests that (even at sites that were spatial maxima in fluvial sediment transport) there were many dry years during the time span of fan accumulation. Fans today have inverted channels on their surfaces, recording dry-climate wind erosion. Perhaps in synch with alluvial fan formation in the low latitudes, Fresh Shallow Valleys formed within latitude bands (25-42)°N and (30-42)°S (Howard & Moore 2011, Mangold 2012, Wilson et al. 2016). The Fresh Shallow Valleys formed over ≫$10^8$ yr, but moved relatively little sediment (Wilson et al. 2016). A younger runoff episode is recorded by uncommon supraglacial or proglacial channels, best seen at Lyot crater < 1 Gya (Dickson et al. 2009, Fassett et al. 2010).

Runoff at Lyot crater – either ice-melt or snowmelt – marks the youngest definite climate-driven water runoff on Mars. Shallow <100°C groundwater at 0.1-1.7 Ga is recorded by meteorites, but it is not clear if this shallow groundwater required a wet climate (Borg & Drake 2005, Swindle 2000, Nemchin et al. 2014). Alluvial fans formed close to (and within) a few mid-sized impact craters ≪1 Ga (e.g. Williams & Malin 2008), but this was caused either by localized precipitation or by dewatering of impact ejecta (Goddard et al. 2014), and not by global climate change. Rovers find evidence for ≪100 Ma surface aqueous alteration (§3). This would require aqueous fluids to infiltrate rocks/regolith, but does not require surface runoff. It is not clear whether or not ≲5 Ma surface features (gullies and Recurring Slope Lineae) record water flow (Dundas et al. 2017a, 2017b; Leask et al. 2018).

In summary, the number of distinct river-forming climates required by the data is ≥2 (e.g. Mangold et al. 2012).

The distinct periods of river-forming climate show different spatial coverage, different river lengths, and different amounts of fluvial sediment transport erosion. These differences are summarized in Fig. 4 and discussed below.

1. The Early/Middle Noachian may be the most erosive time in Mars history – with 1 km of column-averaged sediment transport inferred (Robbins et al. 2013, Irwin et al. 2013). However, rivers were only ~$10^2$ km long, and lakes did not overspill. Lack of overspill implies that climate was arid; that individual wet events did not last long enough to overspill craters; or both. Earth's Holocene average erosion rate was ~$5 \times 10^{-5}$ m/yr (Milliman & Syvitski 1992). If all of the inferred Early/Middle Noachian sediment transport was associated with fluvial erosion, then this is equivalent to ~$2 \times 10^7$ yr of fluvial sediment transport at Holocene Earth rates.
2. The Late Noachian / Early Hesperian valleys represent an intense, but relatively late and topographically superficial, erosion episode. Valleys cut ~$10^2$ m deep, but valleys are wide-spaced (Williams & Phillips 2001, Carr & Malin 2000, Hynek et al. 2010). As a result, dividing valley volume by the area of Noachian terrain (as mapped by Tanaka et al. 2014) yields just 3-4 m areally-averaged erosion (Irwin et al. 2005a, Ansan & Mangold 2013, Luo et al. 2017). This corresponds to only $10^5$ yr of Earth-average fluvial sediment transport. This erosion was not enough for rivers to consistently form smooth convex-up profiles of elevation versus downstream distance (Aharonson et al. 2002, Som et al. 2009, Penido et al. 2013). Many but not all lakes overspilled at this time (e.g. Goudge & Fassett 2018, Fassett & Head 2008b), which was the only time in Mars history for which we see a record of pervasive and regionally-integrated valley networks.
3. The Late Hesperian and Amazonian fluvial erosion is spatially more focused. During this time, low-latitude overspill of large lakes was much less probable than during the Late Noachian / Early Hesperian (Goudge et al. 2016). Overall, planet-median time-averaged erosion in the Late Hesperian and Amazonian was $3 \times 10^{-10}$ – $2 \times 10^{-8}$ m/yr (Golombek et al. 2006, 2014). Nevertheless, some locations (e.g. Fig. 2) saw >1 km erosion - these focused erosion zones covered only a small fraction of the planet. Deep erosion occurs preferentially on slopes that face either N or S (Morgan et al. 2018). Lake (playa?) deposits are seen at the toes of some fans (e.g. Morgan et al. 2014).

In summary, the time spanned by all river-forming climates on Mars was $\gg 10^8$ yr (Fig. 3). But the total sediment moved by rivers corresponds to $\lesssim 10^7$ yr at Earth Holocene rates. This contrast suggests that river-forming climates were intermittent.

### 2.3. Relative to rivers that formed earlier in Mars history, rivers that formed later in Mars history were preferentially at lower elevations (high confidence), and show a stronger dependence on latitude (medium confidence).

The latitude-elevation distribution of rivers and lakes suggests trends over time (Fig. 5).

- A proxy for Early/Middle Noachian crater modification, which included fluvial resurfacing, is the spatial distribution of the Middle Noachian highland and Late Noachian highland units (Fig. 5a) (Tanaka et al. 2014). Those units are found over a wide range of latitudes, and mainly at locally lower elevations (Craddock et al. 1993, Irwin et al. 2013).
- The Late Noachian / Early Hesperian valleys are under-abundant at low elevation, although this may be a preservation artifact (Fig. 5b) (§5.2) (Hynek et al. 2010). After correcting for the availability of non-resurfaced terrain, the data suggest a preference for low latitude (Williams & Phillips 2001).
- In the Hesperian / Early Amazonian (Fig. 5c), the elevation preference is for low elevation. ~¾ of large alluvial fan apices are below 0 m. ~⅘ of light-toned layered sedimentary rock occurrences in the Malin et al. (2010) catalog are below 0 m. Even after correcting for the availability of non-resurfaced terrain, light-toned layered sedimentary rock occurrences in the Malin et al. (2010) catalog are found most frequently at low elevations (Kite et al. 2013).
- Strong latitude preferences are obvious in the Hesperian / Early Amazonian (Fig. 5c). >¾ of large alluvial fans apices are at latitude (30 - 15)°S (Kraal et al. 2008a), and ~⅔ of light-toned layered sedimentary rocks in the Malin et al. (2010) catalog are within 10° of the equator (Kite et al. 2013). Fresh Shallow Valleys are also confined to latitude belts (Mangold 2012, Wilson et al. 2016).

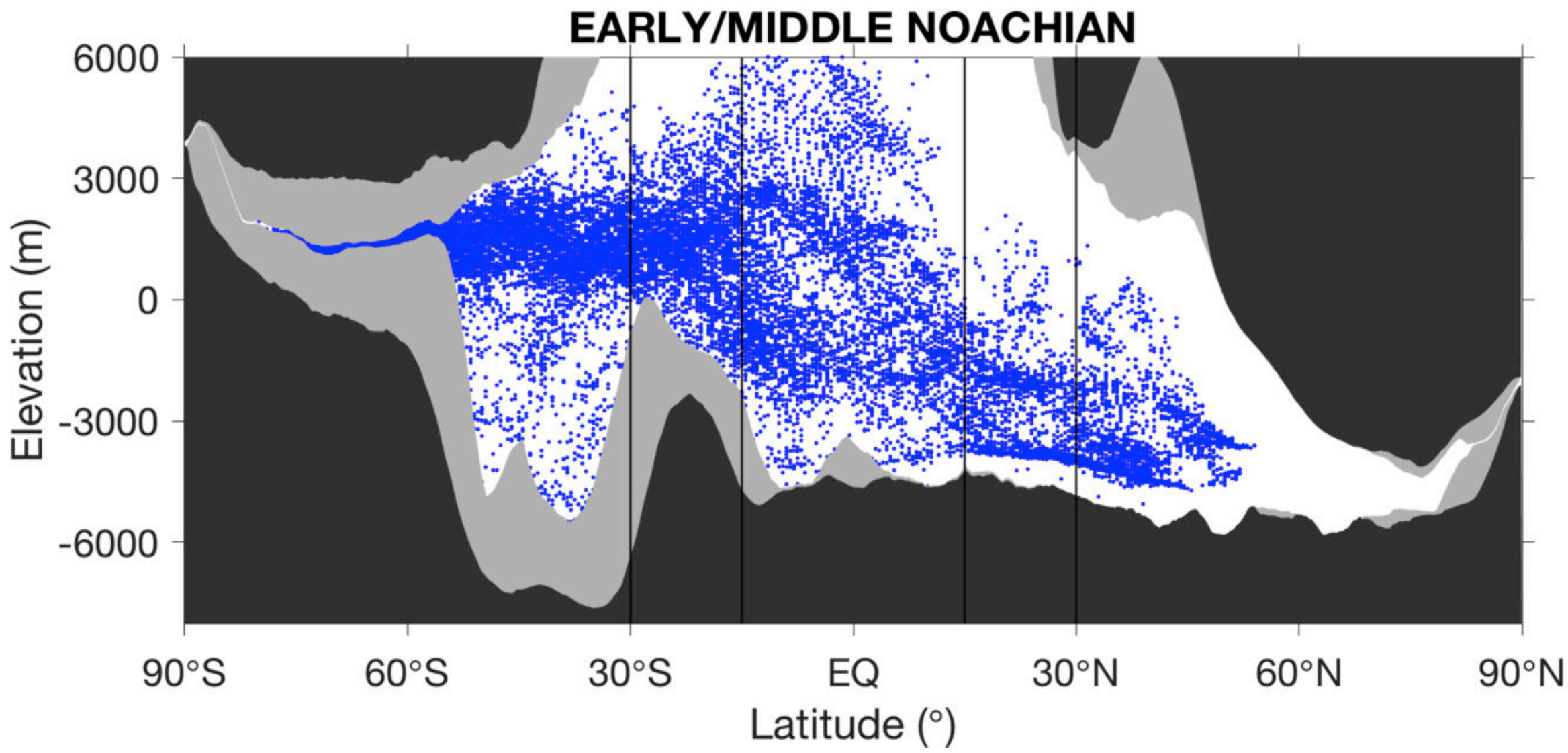
EARLY/MIDDLE NOACHIAN
Elevation (m)
6000
3000
0
-3000
-6000
90°S
60°S
30°S
EQ
30°N
60°N
90°N
Latitude (°)

**(a)**

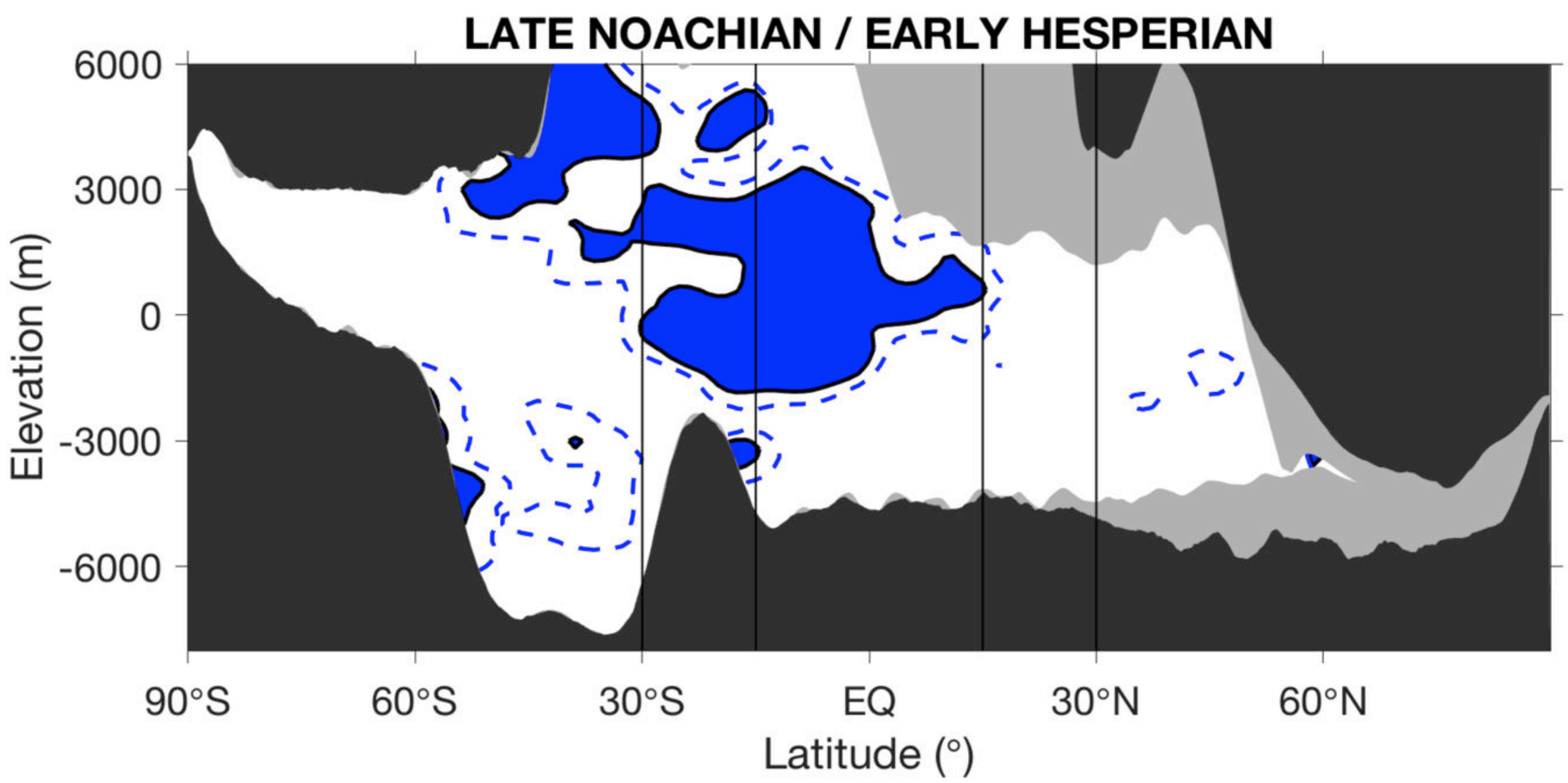
LATE NOACHIAN / EARLY HESPERIAN
Elevation (m)
6000
3000
0
-3000
-6000
90°S
60°S
30°S
EQ
30°N
60°N
Latitude (°)

**(b)**

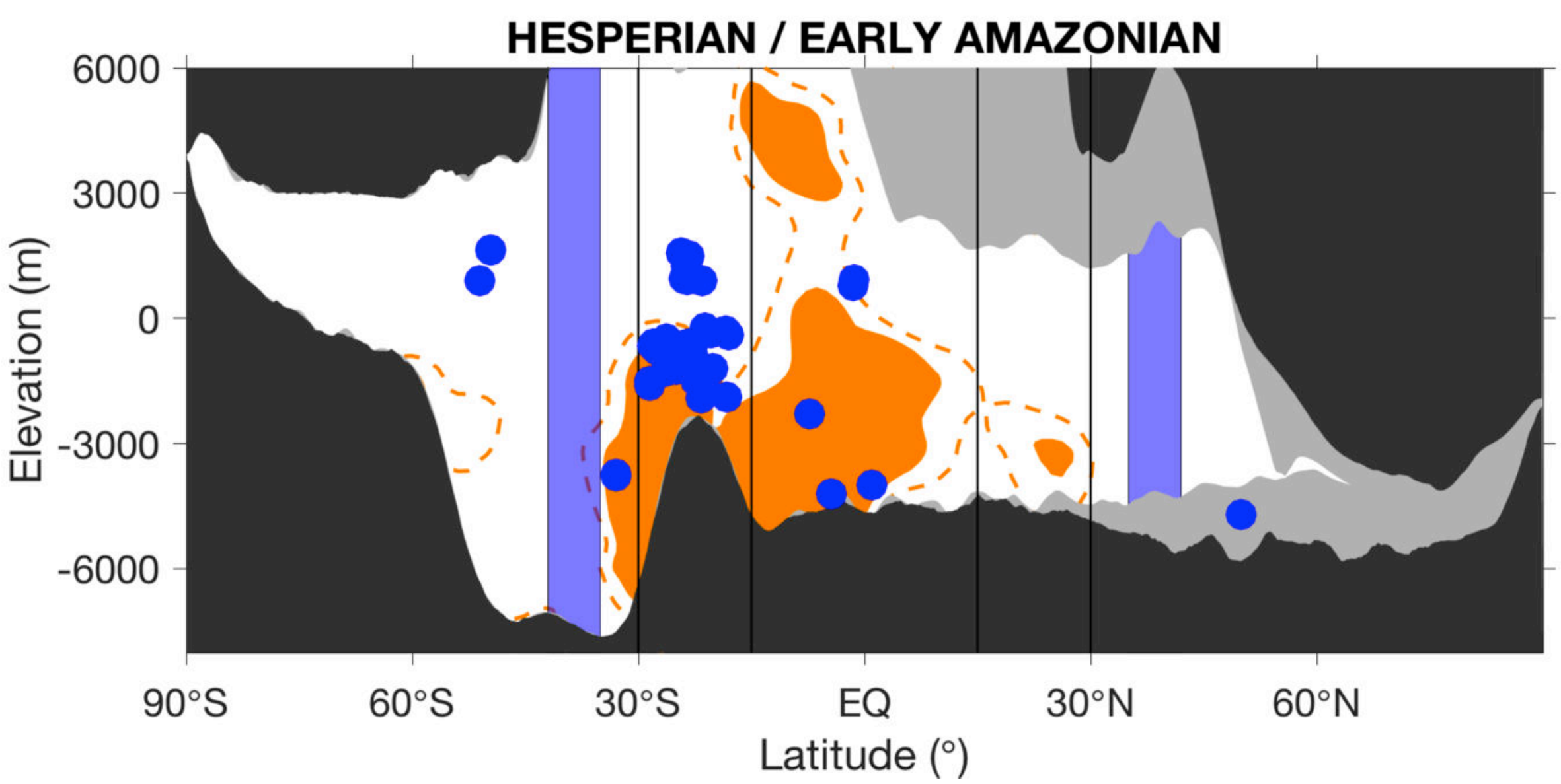
HESPERIAN / EARLY AMAZONIAN
Elevation (m)
6000
3000
0
-3000
-6000
90°S
60°S
30°S
EQ
30°N
60°N
Latitude (°)

**(c)**

**Fig. 5.** Latitude-elevation plots of climate-relevant geologic activity for each period of river-forming climate. The black regions have no data, and the gray regions correspond to terrain that was geologically reset after the time slice in question. Vertical black lines highlight latitudes ±15° and ±30°. Appendix Fig. A1 shows more detail. **(a) Early / Middle Noachian time slice:** Blue dots correspond to Middle Noachian highland materials (Tanaka et al. 2014). These materials are the major Noachian geologic terrains to have been affected by gravity-driven resurfacing (Irwin et al. 2013). **(b) Late Noachian / Early Hesperian time slice:** Blue corresponds to the latitude/elevation zones that contain ⅔ of the valleys (Hynek et al. 2010), after correcting for the nonuniform distribution of latitude/elevation (Appendix Fig. A1). Blue dashed line is the same, but for 9/10 of the valleys. **(c) Late Hesperian / Amazonian time slice:** Blue disks mark large alluvial fans (combining catalogs of Moore & Howard 2005 and Kraal et al. 2008a). Pale blue stripes mark latitude range of Fresh Shallow Valleys (Wilson et al. 2016). Orange shading corresponds to the latitude/elevation zones that contain ⅔ of the light-toned layered sedimentary rocks (Malin et al. 2010), after correcting for the nonuniform distribution of latitude/elevation (Appendix Fig. A1). Orange dashed line is the same, but for 9/10 of the light-toned layered sedimentary rocks.

To sum up, Fig. 5 suggests a shift from early control by elevation on fluvial sediment transport, to later control by latitude. These trends are consistent with theoretical expectations for atmospheric decay (Wordsworth et al. 2013, Wordsworth 2016). Theory predicts that $\gtrsim$ 0.3 bar atmospheric pressure would cause surface temperature – and thus the potential for above-freezing temperatures needed for snowmelt and/or rainfall – to be controlled by elevation (Wordsworth 2016). Theory predicts that as pressure dropped below ~0.1 bar, surface temperature would become more sensitive to direct insolation, which is a function of latitude (Kite et al. 2013). Combining data and theory hints that – during the Hesperian – the atmospheric pressure fell from >0.3 bar to <0.1 bar.

### 2.4. *τ-R-I-N* framework

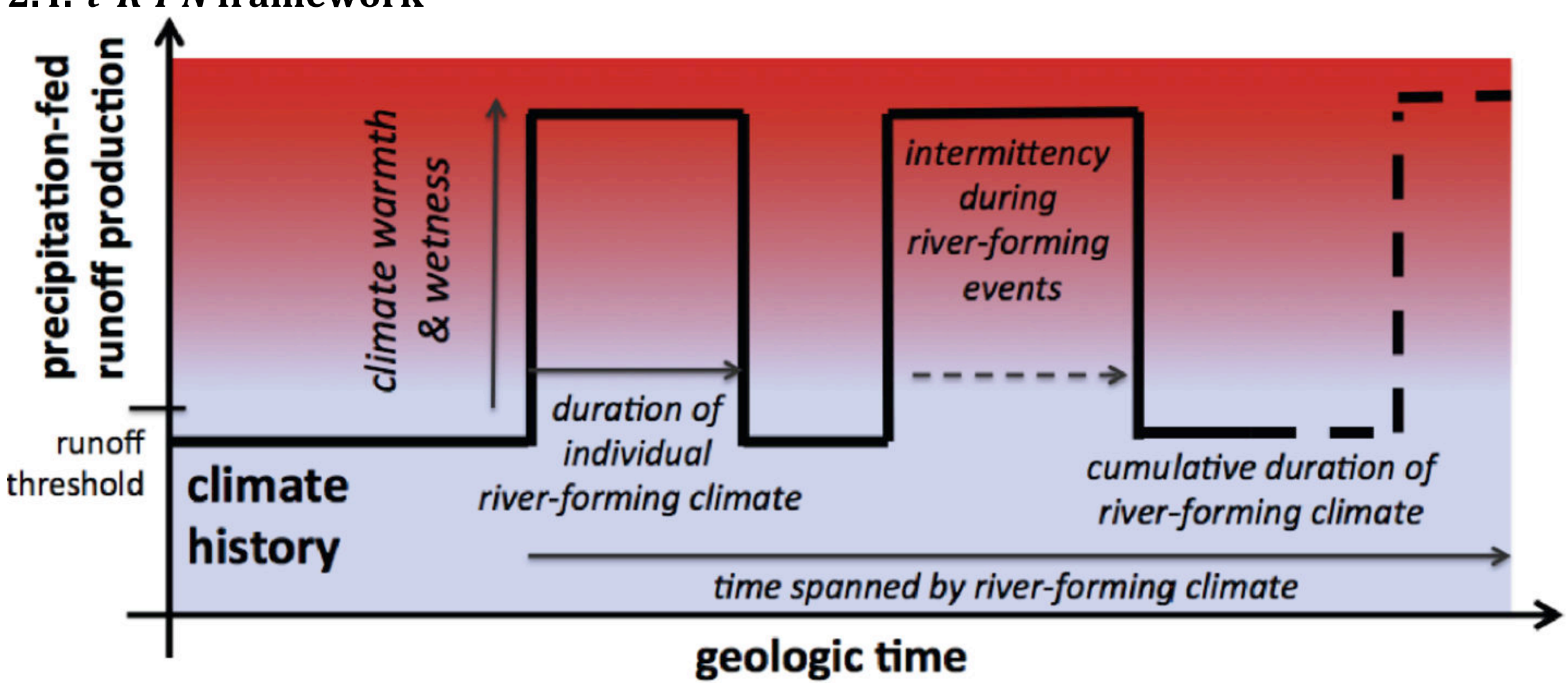

**Fig. 6.** Anatomy of a single river-forming period on Mars.

For each river-forming period, we can sum up the climate using four numbers (Fig. 6). These are:

***τ*** **– Duration of wet climate (yr).** How long did the wet climate persist (without a drought that lasted centuries or longer)?

***R*** **– Peak runoff production (discharge / area, units mm/hr).** How wet was it during the wettest day of the year? *R* is related to climate warmth, because warm conditions are needed for rain and storms.

***I*** **– Intermittency (peak runoff production × hours in year / annual runoff; dimensionless).** Relative to peak runoff, how wet was it on average during the wet years? This is closely related to the "flashiness" and seasonality of the wet climate.

***N*** **– Number of wet events during climate episode (e.g., number of orbital peaks).** A single runoff-producing period may consist of alternations between periods with some runoff in most years, and periods with many consecutive years of zero runoff. For example, wet-dry cycles may be paced by orbital variations (Metz et al. 2009a). How often were the cycles of wet and dry repeated?

Constraints for these *τ*-*R*-*I*-*N* parameters are discussed below (§2.5-§2.8).

The time spanned by river deposits at a given site is also of interest (Fig. 6). This is constrained for one site of unknown age to ≳(1-20) Myr (Kite et al. 2013b).

## 2.5. τ: Duration of longest individual river-forming climate: >($10^2$ – $10^3$) yr (medium confidence), >2 × $10^4$ yr (low/medium confidence).

*Medium confidence – orbiter data:* At least some Hesperian closed-basin lakes lasted >($10^2$ - $10^3$) yr, based on hydrologic analysis (Irwin et al. 2015, Williams & Weitz 2014). The analysis starts from measurements of delta volume. (Deltas existed at Eberswalde and Holden, and evidence has been reported for deltas at many other sites – e.g. Lewis & Aharonson 2006, Grant et al. 2008, Metz et al. 2009b, di Achille & Hynek 2010, Goudge et al. 2017, Cardenas et al. 2018, Goudge et al. 2018, Adler et al. 2019). Multiplying delta volume by assumed water:sediment volume ratio >$10^2$ gives (at Eberswalde, SW Melas, and Gale) a cumulative water volume that is far in excess of the paleo-lake volume. This volume excess is inconsistent with the roughly constant lake level recorded by the un-incised delta deposits. Therefore, as the excess water was supplied, it must have been removed. The water removal processes (for closed-basin lakes, for which there is no outlet channel) are infiltration and evaporation. Both these water-removal processes are slow (Irwin et al. 2015). Therefore, water supply must also have been slow, and delta build-out rate would have been slow as well. Dividing delta volume by these slow delta build-out rates gives >($10^2$ – $10^3$) yr for the delta at Eberswalde (Irwin et al. 2015, Kite et al. 2017a) (Fig. 7). The main uncertainty is the water:sediment ratio (e.g. Kleinhans 2005, Kleinhans et al. 2010, Mangold et al. 2012). In contrast to the long life inferred for the delta at Eberswalde, formation by brief, low water:sediment ratio floods is inferred for many other sites that have deltas which show a stepped morphology. This stepped morphology results (in laboratory experiments) from "one-shot" delivery of water and sediment, concurrent with rapid lake-level rise (e.g. de Villiers et al. 2013, Hauber et al. 2013).

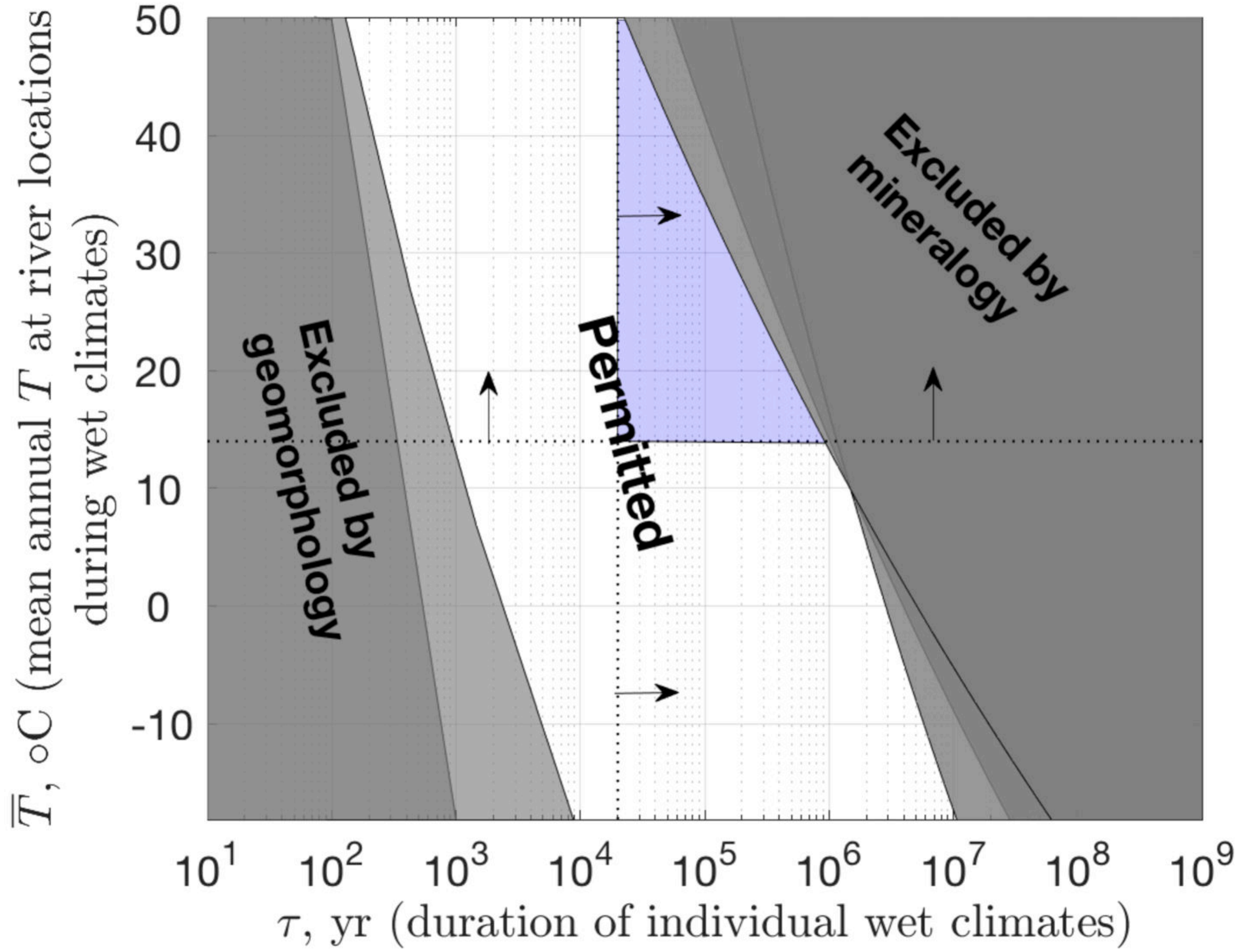

**Fig. 7.** Constraints on the temperature and timescale of post-Noachian river-forming climates on Mars. The blue-tinted region is not disfavored by any geologic data. The geomorphic constraint boundaries are for water:sediment volume ratio = $10^2$. These boundaries are at smaller durations at higher temperatures because of the *T*-dependence of evaporation (Irwin et al. 2015). The less permissive geomorphic boundary slope is drawn using the equation of Eagleman (1967) for a relative humidity of 50%. The more permissive geomorphic boundary slope is adjusted to pass through the evaporation rate 0.8 m/yr at -18°C (Dugan et al. 2013) and an evaporation rate of 10 m/yr at 50°C. The mineralogy constraint boundary is at smaller durations for higher temperatures because of the *T*-dependence of olivine dissolution (Olsen & Rimstidt 2007). The three different mineralogy constraint boundaries correspond to the full range (42 kJ/mol, and 79.5 kJ/mol) and average of (63 kJ/mol) of activation energies for olivine dissolution reported by Olsen & Rimstidt (2007). The dotted lines correspond to lower confidence constraints. The horizontal dotted line is a lower bound on temperature from ALH 84001 carbonate $\Delta_{47}$ (Halevy et al. 2011). The vertical dotted line is a lower bound on lake lifetime from *Curiosity* stratigraphic logs (§2.5).

During the Late Noachian / Early Hesperian valley-network-forming period, hundreds of exit-breach lakes existed (Fassett et al. 2008, Grant et al. 2011). In order to breach their rims, these lakes had to fill up. Filling up with precipitation-sourced water would have required $>10^2$ yr; thousands of years in the case of the Eridania sea (Irwin et al. 2004).

Both the lake hydrology lower limit and the lake fill-up duration lower limit are lower limits on the duration of the corresponding river-forming climates.

*Low/medium confidence - rover data:* The Mars Science Laboratory *Curiosity* rover sampled a mudstone in Gale crater – the Murray – that has rhythmic mm-scale laminations (Grotzinger et al. 2015, Hurowitz et al. 2017). If these laminae are interpreted as annual lake varves, then the 25 m – 30 m thickness of stratigraphic sections with mm-thick laminae, and without textural evidence for drying-up (Fedo et al. 2017, Stein et al. 2018), suggests lake lifetime $>2 \times 10^4$ yr. However, textural evidence for drying-up is observed in the best-studied part of the Murray (Kah et al. 2018), and periodic drying-up might have occurred without leaving textural evidence (Bristow et al. 2018).

The medium confidence $>(10^2 - 10^3)$ yr limit is inconsistent with wet climates caused by the direct conversion of impact kinetic energy into water vapor. Post-impact rainout would last < 3 yr (Steakley et al. 2017, Turbet et al. 2017a). The $>(10^2 - 10^3)$ yr limit is also inconsistent with wet climates maintained by single, isolated super-volcanic eruptions. Such eruptions would last < 30 yr (Halevy & Head 2014, Kerber et al. 2015).

### 2.6. *R:* Peak runoff production >0.1 mm/hr (high confidence).

Runoff production for river channels draining catchments of known area may be estimated using channel slope, width, and depth. Channel width and slope can be measured from digital terrain models and anaglyphs, exploiting Mars' tectonic quiescence [3] and preservation of channel properties in river deposits (e.g. Hajek & Wolinsky 2012). Channel depth can be estimated for self-formed channels using rover measurements of the grainsize of river-bed material. That is because the channel depth adjusts to move river-bed material downstream. Unfortunately the four data needed to back out runoff production from first principles – drainage area, plus channel width, slope, and depth – are almost never simultaneously available for Mars rivers. There are two possible work-arounds: (1) Use empirical correlations from Earth rivers, corrected for Mars gravity (Parker et al. 2007, Eaton et al. 2013, Li et al. 2015, Pfeiffer et al. 2017). (2) Guess a grain size based on Earth analogy, or observations made at or beyond the pixel scale of orbiter images (e.g. Morgan et al. 2014). For any individual measurement, both methods have order-of-magnitude uncertainty (Jacobsen & Burr 2016, Dietrich et al. 2017). Nevertheless, both approaches consistently indicate runoff production > 0.1 mm/hr on Early Mars (Palucis et al. 2014, Fassett & Head 2005, Irwin et al. 2005b, Jaumann et al. 2005,

---

[3] Although tilting by planetary tectonics has little effect on Mars river slopes, within some sedimentary basins differential compaction and subsidence has tilted layers substantially (e.g. Lefort et al. 2012, Gabasova & Kite 2018).

Baratti et al. 2015, Dietrich et al. 2017), with >1 mm/hr suggested at many locations (Kite et al. 2018a). This is a conservative lower limit on peak melt/rain rates, for two reasons. (1) Storm runoff peaks are damped as they flow downstream (Dingman 2014). (2) Climate models must produce rain/melt rates in excess not just of runoff production, but rather of (runoff + infiltration + refreezing-within-snowpack + evaporation) (Clow 1987).

Precipitation-fed runoff production is a probe of past Mars temperature. When the air is warm, snow melts faster; when the air is warmer still, rain and storms may occur.

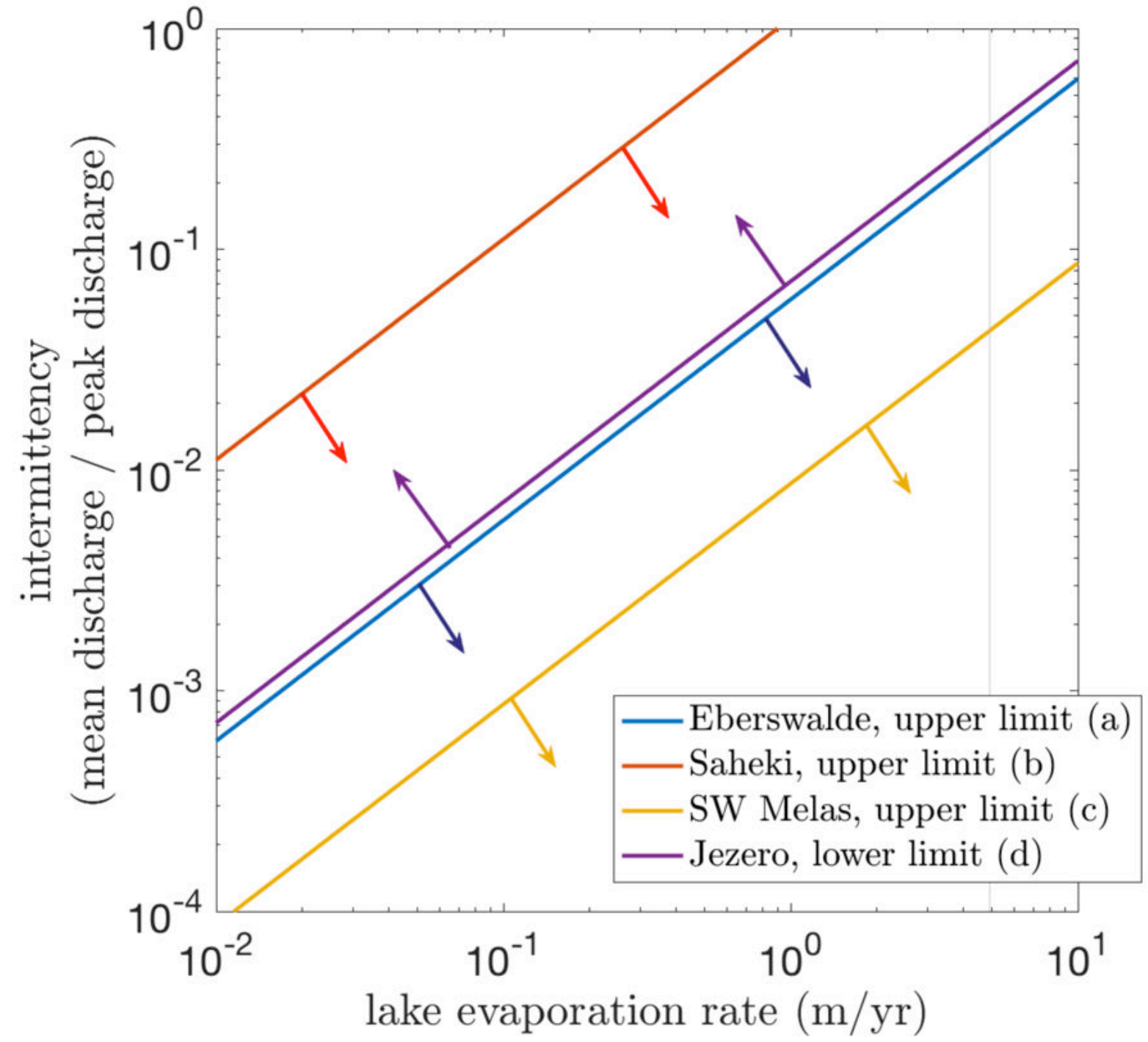

**Fig. 8.** Intermittency constraints for Early Mars river-forming climates. Example sites: (a) Constant lake level at Eberswalde crater. $Q_{water}$ = 400 m$^3$ s$^{-1}$, $A_{lake}$ = 710 km$^2$ (Irwin et al. 2015). (b) Non-formation of a deep lake at Saheki crater. $Q_{water}$ = 30 m$^3$ s$^{-1}$ (Morgan et al. 2014), $A_{playa}$ < 10$^3$ km$^2$. (c) Constant lake level at SW Melas Chasma. $Q_{water}$ = 250 m$^3$ s$^{-1}$, $A_{lake}$ = 65 km$^2$ (Williams & Weitz 2014). (d) Overspill of the lake at Jezero crater. $Q_{water}$ = 700 m$^3$ s$^{-1}$ (Fassett & Head 2005), $A_{lake}$ = 1500 km$^2$. The gray vertical line at ~6 m/yr marks an energetic upper limit (400 W/m$^2$) for Early Mars climate-driven evaporation rate. For sites (a)-(c), the main assumption is that the climate event lasted long enough to fill up the lake at continuous flow (e.g., >7 years for Eberswalde; Irwin et al. 2015). The non-agreement between the Jezero constraint and the Saheki constraint implies that river-forming climates varied in time, space, or both (Goudge et al. 2016).

### 2.7. *I:* Intermittency during wet events: peak runoff production <10% of the time (high confidence).

Even during the wettest period in Mars history, not all lakes overflowed. To avoid all lakes overflowing, given $R$ > 0.1 mm/hr and $\tau$ > (10$^2$-10$^3$) yr, runoff production during the wet events must have been intermittent (Fig. 8) (e.g. Barnhart et al. 2009, Matsubara et al. 2011). As on Earth, intermittency might correspond to seasonality, or to infrequent storms (Hoke et al. 2011). Intermittency on longer timescales is suggested by extremely slow alluvial fan build-up rate averaged over the Late Hesperian / Amazonian wet event (Kite et al. 2017). This suggests that wet years were intercalated with years too dry for river runoff (Buhler et al. 2014). Years too dry for river runoff need not be too dry for life; life persists in climates too dry for rivers (Amundson et al. 2012). Moreover, it is possible that alluvial fan activity was not globally synchronous. Therefore, the at-a-site runoff-production intermittency is a lower bound on the global intermittency. The main uncertainty in estimating intermittency is lake evaporation rate (Fig. 8). Taking into account an energetic upper limit on lake evaporation rate, $I$ < 10% for the rivers that delivered sediment to closed-basin sites (such as SW Melas) during the Late Hesperian / Amazonian.

### 2.8. $N \times \tau$ = cumulative wet years during valley-network-forming climate episode: >$10^5$ yr (medium confidence).

Around the Noachian/Hesperian boundary, ~$2 \times 10^5$ $km^3$ sediment moved downstream to form Mars' valley networks (Luo et al. 2017). Dividing the volume of sediment by the sediment flux gives a timescale >$10^4$ yr to form the valley networks (Hoke et al. 2011, Orofino et al. 2018, Rosenberg et al. 2019). This increases to >$10^5$ yr if intermittency is considered. The sediment flux is estimated using the relation between sediment flux and water discharge (e.g. Marcelo Garcia 2008, Parker et al. 2007). In turn, water discharge is estimated using channel width and depth. These sediment flux calculations assume that the Mars rivers that cut the valley networks were gravel-bedded. This is probably a safe assumption. Where measured, in-place Mars clastic rock strength is similar to adobe bricks or weak concrete, not to loose sand (Thomson et al. 2013, Peters et al. 2018). Although shorter valley formation durations are possible if the Mars valley networks were cut into loose, fine-grained sediment (Rosenberg & Head 2015), that set-up does not match in-place measurements of Mars rock strength.

Published THEMIS crater counts do not exclude the hypothesis that all of the regionally-integrated valley networks on the Southern highlands were incised in one brief interval (Fassett & Head 2008, Warner et al. 2015; but see also Hoke & Hynek 2009). This one-pulse hypothesis for valley-network formation might be tested in future by seeking craters interbedded during the valley networks (Hartmann 1974).

### 2.9. Maximum lake size: >$1.1 \times 10^6$ $km^2$ (very high confidence).

*Very high confidence:* A spillway drained a >$1.1 \times 10^6$ $km^2$ sea in Eridania during the Late Noachian (Irwin et al. 2004). This Eridania sea would have required at least thousands of years to fill to the >1 km depths needed to over-top the spillway. Seas with volume >$10^5$ $km^3$ existed within Valles Marineris (very high confidence) (Warner et al. 2013, Harrison & Chapman 2008). It is not certain that the Valles Marineris lakes were precipitation-fed.

*Low confidence:* Bigger paleo-seas have been proposed. For example, the Northern Ocean hypothesis is that a >$1.2 \times 10^7$ $km^2$ sea existed in the northern lowlands of Mars (Parker et al. 1993, Clifford & Parker 2001, di Achille & Hynek 2010, Ivanov et al. 2017, Citron et al. 2018). Indeed, if Mars formed with surface $H_2O$, that water must have drained into the northern lowlands, at least in the immediate aftermath of Hellas-sized impacts (Early Noachian). However, the Parker et al. (1993) hypothesis posits an ocean in the Late Noachian, or even later. In favor of this hypothesis, it is plausible that Late Noachian – Early Amazonian lakes and rivers in the highlands of Mars should have a counterpart in the lowlands. Moreover, a candidate shoreline exists close to an equipotential (Ivanov et al. 2017). In opposition, high-resolution images of the candidate shoreline are ambiguous (Malin & Edgett 1999, Carr & Head 2003, Ghatan & Zimbelman 2006, Salvatore & Christensen 2014). Hesperian deposits claimed to record a past ocean tsunami (Rodriguez et al. 2016) might instead be wet ejecta (Grant & Wilson 2018) from the impact that formed Lyot crater. Hundreds of delta candidates exist on Mars (di Achille & Hynek 2010), but few deltas have been definitively identified from orbit. Definitive delta identification requires good outcrop exposure, and analysis of highest-resolution Digital Terrain Models. Of the deltas on Mars for which the strongest evidence exists (e.g. Malin & Edgett 2003, Grant et al. 2008, diBiase et al. 2013, Goudge et al. 2017, Hughes et al. 2019), none drain into basins that are topographically open to the northern lowlands based on modern topography. Overall, the Parker et al. (1993) hypothesis is not proven.

A Northern Ocean (but not an Eridania-sized sea) would kick-start planet-scale albedo feedbacks and water-vapor feedbacks that might help to warm Mars (Mischna et al. 2013, but see also Turbet et al. 2017b).

## 3. Climates that allowed sedimentary rocks, shallow diagenesis, and soil weathering.

Ancient sedimentary materials are indurated and strong – "sedimentary rocks" (Malin & Edgett 2000). Rock-making probably involves cementation by aqueous minerals (Gendrin et al. 2005, Murchie et al. 2009, Grotzinger & Milliken 2012, McLennan et al. 2019). Cementation requires transport of ions by some aqueous fluid (McLennan & Grotzinger 2008, Hurowitz & Fischer 2014, but see also Niles et al. 2017). However, the corresponding aqueous fluid supply rate requirement is much lower than for the river-forming climates (e.g. Arvidson et al. 2010). In particular, the aqueous fluids can be supplied by either quickly by rain/snowmelt, or slowly by groundwater (Andrews-Hanna et al. 2010). Because groundwater upwelling can occur in a cold climate, it is not as challenging to explain sedimentary rock formation as it is to explain the river-forming climates. Indeed, acidic or saline fluids can persist at temperatures well below 0 °C (Fairén et al. 2009, Ward & Pollard 2018) – one Antarctic pond stays unfrozen to -50 °C (Toner et al. 2017), and the $H_2O$-$H_2SO_4$ eutectic is at -74 °C (Niles et al. 2017, Niles & Michalski 2009). With these caveats, the low-rate aqueous fluid supply might constrain climate models, if its duration were known.

The first sedimentary-rock basins to be catalogued from orbit were light-toned, layered sedimentary rocks (Malin & Edgett 2000). Although these materials mostly postdate the valley networks, the depositional record of later river-forming climates is intercalated within the sedimentary rock record (Kite et al. 2015, Grotzinger et al. 2005, Milliken et al. 2014, Williams et al. 2018). The light-toned, layered sedimentary rocks that have been catalogued from orbit cover <3% of planet surface area (Tanaka et al. 2014). However, this is only a subset of the true distribution of sedimentary rocks on Mars, which is broader in latitude and age (Malin 1976, Edgett & Malin 2002, McLennan et al. 2019). For example, the northwest part of Gale crater's moat appears bland from orbit, but is a smorgasbord of sedimentary rocks and aqueous weathering at rover scale (e.g. Stack et al. 2016, Buz et al. 2017).

### 3.1. Years of sediment deposition recorded by sedimentary rocks: >$10^7$ yr (high confidence).

Many of the sedimentary rocks show rhythmic bedding (Lewis et al. 2008, Lewis & Aharonson 2014). Rhythmic bedding can be tied to orbitally-paced variations in insolation (at the equator: $2.5 \times 10^4$ yr – $1.2 \times 10^5$ yr period), allowing the calculation of sediment build-up rate. Sediment build-up rate so determined is $10^{-5}$-$10^{-4}$ m/yr. This build-up rate, combined with the total thickness of rhythmite-containing rock, gives build-up duration (Lewis & Aharonson 2014). The resulting duration is >$10^7$ yr. During this >$10^7$ yr period, liquid water does not have to be available every year. To the contrary, the sedimentary rock record could be a "wet-pass filter" of Mars history, only recording those times that produced liquid water (Kite et al. 2013). If the "wet-pass filter" idea is correct, then long dry periods would correspond to hiatuses or wind-erosion surfaces (unconformities) (Banham et al. 2018). Indeed, one of the erosion surfaces within sedimentary sequences has a crater count corresponding to erosion/nondeposition duration >$10^8$ yr (Kite et al. 2015).

The total time spanned by sedimentary-rock build-up was $\gg 10^8$ yr, based on crater counts and crosscutting relationships (Middle Noachian to Amazonian; Kerber & Head 2010, Hynek & Di Achille 2017).

### 3.2. Water column required to lithify sedimentary rocks: >20 km (medium/high confidence).

The thickest known sedimentary rock accumulation on Mars (the easternmost mound within East Candor Chasma; 8°S 66°W) is >8 km thick. It is reasonable to assume that this rock column has approximately the same composition as that of the light-toned sedimentary rocks investigated by Mars Exploration Rover *Opportunity* at Meridiani (5.5 wt% $H_2O$; Glotch et al. 2006, Bibring et al. 2007). With this assumption, ~1 km column $H_2O$ is bound within the rock as $H_2O$ or as OH (Wang et al. 2016). This mineral-bound water is less than the amount of water involved in cementing these rocks. The needed water is estimated using geochemical reaction models. These models yield estimated water:rock mass ratios of $\gtrsim 1$. The water:rock mass ratio in geochemistry refers to the time-integrated amount of water that participates in reactions with the rock (Reed 1997). This geochemical ratio is distinct from the water:sediment ratio in geomorphology. For example, suppose a flood moves olivine pebbles downstream, and the flood water then quickly evaporates. In this case, the geomorphologist's water:sediment ratio is high. However, the geochemist's water:rock ratio need not be high, because – in this case – the olivine does not have enough time to dissolve. As another example, suppose rain water trickles down through a soil profile for Myr. Then, the instantaneous volume ratio of water to sediment is no more than ~0.4 (limited by soil porosity). But the time-integrated water:rock mass ratio in this set-up can be >$10^4$.

Water:rock mass ratio ≳1 yields water column >20 km. This water could be supplied either by precipitation (Kite et al. 2013) or by deep-sourced groundwater (Andrews-Hanna et al. 2010) [4]. Groundwater, either deep-sourced or locally derived, cemented the sedimentary rocks at shallow depths (McLennan et al. 2005, Siebach et al. 2014, Martin et al. 2017), and groundwater transport through fractures occurred after the sedimentary rocks were buried deep enough to be lithified (e.g. Okubo & McEwen 2007, Vaniman et al. 2018). Groundwater circulation could not have been both globally pervasive and long-lasting however, because salts (and other easily-reset minerals in the rocks) have remained in place since the Early Hesperian.

**3.3. Duration of post-Noachian surface liquid water at an "average" place on Mars: <$10^7$ yr (medium/high confidence).**

Olivine in liquid water dissolves quickly (<$10^7$ yr, usually ≪$10^7$ yr). Despite this fragility, olivine is widespread in Mars rocks and soil (Hamilton & Christensen 2005, Koeppen & Hamilton 2008, McGlynn et al. 2012, Ody et al. 2013). Olivine's persistence – alongside other minerals which dissolve or transform readily in water, such as halite, jarosite, and amorphous phases including hydrated amorphous silica – shows that post-Noachian aqueous alteration on Mars could not have been both global and long-lasting (e.g. Osterloo et al. 2010, Stopar et al. 2006, Tosca & Knoll 2009, Elwood-Madden et al. 2009, Ruff & Hamilton 2017, Tosca et al. 2008, Olsen & Rimstidt 2007, McLennan et al. 2019). To the contrary, either liquid water was available briefly, or patchily, or both. Although physical erosion can "reset" the mineral-dissolution stopwatch (e.g., Hausrath et al. 2008), this does not explain the persistence of post-Noachian olivine at places that show little evidence for physical erosion. At the *Opportunity* landing site, Fe is mobile but Al is not mobile, suggesting water-limited and time-limited element mobility at water-to-rock mass ratio < 300 (Hurowitz & McLennan 2007) and with water available for <20 years per sediment parcel (Berger et al. 2009). These quantitative upper limits are consistent with qualitative inferences about aqueous alteration. In particular, much of the climate-driven surface aqueous alteration uncovered by rovers is cation-conservative (e.g. Ehlmann et al. 2011a, Thompson et al. 2016), although there are open-system exceptions (e.g. Michalski et al. 2015). Cation-conservative alteration is in contrast to Earth and suggests cold conditions with limited liquid water (Milliken & Bish 2010, Ehlmann et al. 2011a). Low water:rock ratio is also inferred at regional scale. At ≥300 km scale, soluble K has not separated from insoluble Th (Taylor et al. 2006). Moreover, aqueous alteration younger than 3.5 Ga is minor on Mars (e.g. Yen et al. 2005, Salvatore et al. 2014). For example, surface wetting ≲$10^8$ yr is limited to surface salt precipitation and near-surface vertical transport of salts (Knoll et al. 2008,

[4] Much more water is indicated by the acid-titration calculations of Hurowitz et al. (2010). Applied to the 8°S 66°W site, they give water columns of 2,000 km (for pH = 2) or 200,000 km (for pH = 4).

Wang et al. 2006, Arvidson et al. 2010, Squyres et al. 2009, Amundson 2018). A largely dry history is also recorded by the soil and dust (Goetz et al. 2005, Yen et al. 2005, Pike et al. 2011, Cousin et al. 2015).

In summary, climate-driven surface aqueous alteration on Mars had shut down in most places by ~3.4 Ga (Ehlmann et al. 2011a, Edwards & Ehlmann 2014). During the post-3.4 Ga river-forming climate (>1 km column runoff production) that produced the alluvial fans, much of the planet's surface apparently evaded aqueous alteration. This contrast between the evidence for alluvial fans, lakes and rivers (§2 of this paper), and mineralogical/chemical evidence for low water-rock ratios (§3 of this paper), has been described as "[a] fundamental paradox" (McLennan 2012). To resolve this paradox will require climate modelers to consider one or more of low temperatures, intermittency, patchiness, strong positive feedbacks, or bistability (Head 2017, Ehlmann et al. 2016, Ehlmann & Edwards 2014, McLennan 2012).

**3.4. Example rover observations of ancient sediments.**

Surface conditions inferred from rover data for Early Mars are consistent with slow water-limited alteration interspersed with $<10^6$-yr-long pulses of abundant water.

For example, *Opportunity* at Meridiani Planum found rocks derived from siliciclastic evaporites, that had been reworked by wind and deposited and cemented in association with a shallow, fluctuating groundwater table (McLennan et al. 2005, Grotzinger & McLennan 2008). The persistence of the easily-dissolved mineral olivine, combined with the lack of evidence for Al mobility, indicates rapid formation with little exposure to liquid water after deposition (Hurowitz & McLennan 2007, McLennan 2012, Berger et al. 2009, Elwood Madden et al. 2009).

Mars Exploration Rover *Spirit* was sent to Gusev, which is a crater downstream along the flow-path carved by the draining of the Eridania sea. *Spirit* found water-altered deposits (impact ejecta and wind-transported material) that appear to drape hills (Ming et al. 2006, Squyres et al. 2006, McCoy et al. 2008, Crumpler et al. 2011). These rocks have been interpreted to record two endmember styles of top-down weathering. The first endmember style involves water-limited weathering at rates slower than exist anywhere on today's Earth (Ruff & Hamilton 2017). The second endmember style involves rapid alteration by brief pulses of water that deposited abundant salts (sulfates in some rocks, carbonates in others), but that did not last long enough to dissolve all of the olivine (Squyres et al. 2006, Ruff et al. 2014).

Most rocks sampled by *Curiosity* at Gale crater show >15 wt % clays (Bristow et al. 2018). At Gale, persistence of ferrihydrite shows that the rocks were not buried deeply, nor permeated by hydrothermal fluids (Dehouck et al. 2017, Borlina et al. 2015). Rocks sampled early in the mission show near-isochemical alteration (McLennan et al. 2014, Thompson et al. 2016, Siebach et al. 2017), and rocks sampled later in the mission show evidence consistent with open-system alteration (Mangold et al. 2019, Bristow et al. 2018).

The high abundance of amorphous phases is consistent with cold conditions, low water-rock ratio diagenesis, or both.

All three rovers found that at least some early Mars sedimentary rocks experienced acid alteration (Hurowitz & McLennan 2007, Hurowitz et al. 2006, Ming 2006, Hurowitz et al. 2010, Berger et al. 2017, Rampe et al. 2017).

## 4. Atmospheric pressure and surface temperature.

### 4.1. Paleo-atmospheric pressure, *P*: 0.012-1 bar (low/medium confidence).

There are few firm constraints on Mars paleo-atmospheric pressure (Fig. 9). Today's Mars atmosphere is 95% $CO_2$ (Webster et al. 2013). The modern atmosphere+(ice cap) $CO_2$ reservoir is 0.012 bar (Bierson et al. 2016). That $CO_2$ reservoir was probably larger in the past: the spacecraft-era $CO_2$ escape-to-space flux is ≤0.02 bars/Gyr (Jakosky et al. 2018), but $\gtrsim 10^{-4}$ bars/Gyr (Barabash et al. 2007). Past escape rates were higher, because the drivers of escape (solar wind and solar UV) had greater flux in the past. Modelers of past escape can use spacecraft-era measurements for calibration (Lundin et al. 2013, Lillis et al. 2015, Ramstad et al. 2018). These calibrated models output that post-Hellas atmospheric escape-to-space "would have been as much as 0.8 bar $CO_2$ or 23 m global equivalent layer of $H_2O$" (Jakosky et al. 2018; Lillis et al. 2017). However, because the partitioning between $CO_2$ and $H_2O$ escape is not known, the lower bound on the post-3.4 Ga $CO_2$ escape-to-space flux remains small: $\gtrsim 10^{-3}$ bar (Fig. 9). Moreover, escape-to-space measurements do not constrain sequestration of $CO_2$ in the subsurface, as carbonate or as $CO_2$-clathrate (Wray et al. 2016, Kurahashi-Nakamura & Tajika 2006, Longhi 2006). Finally, escape-to-space models constrain the atmosphere+(ice cap) $CO_2$ reservoir, but what matters for climate is atmospheric $CO_2$. Therefore, in order to reconstruct Mars' atmospheric-pressure history, we need both continued observations/analyses of Mars atmospheric escape (Jakosky et al. 2018, Lee et al. 2018) and also paleo-proxy data.

After ~4.0 Ga, geologic proxy data constrain atmospheric pressure. Atmospheric pressure ~(3.6-3.7) Ga is constrained by river deposits that cocoon small impact craters. The presence of small hypervelocity impact craters within the river deposits shows that the atmosphere was thin around the time the rivers were flowing, because thick atmospheres would slow down and/or disintegrate small impactors (Vasavada et al. 1993). These small-ancient-crater observations give $P$ < (1-2) bar according to one bolide burn--up/break-up model (Kite et al. 2014, Williams et al. 2014). However, this might correspond to periods of atmospheric collapse interspersed with river-forming climates (Soto et al. 2015). $P$ ~ 0.01 bar suggested by ~3.6 Ga bedforms (Lapotre et al. 2016) might also record times of atmospheric collapse. Claims for low $pCO_2$ in the Hesperian based on the nondetection of carbonates in ancient sediments (Bristow et al. 2017) are not supported by experiments (Tosca et al. 2018, Gaudin et al. 2018). Given that higher-than-modern atmospheric pressure is considered essential to river-forming climates on Mars (Hecht 2002, Wordsworth 2016), there is remarkably little direct geological evidence for instantaneous $P$ > 0.012 bar. An exception is the Littleton meteorite at Gale crater. This meteorite is intact, but would have blown up in the atmosphere unless $P$ > (0.012-0.044) bar (Chappelow et al. 2016). A single volcanic bomb sag observed by the Spirit rover gently deflected underlying layers (~3 Ga?), suggesting $P > 0.12$ bar (Manga et al. 2012).

Meteorite noble gas data have been interpreted to require $P > 0.5$ bar at ~4.1 Ga, but also $P < 0.4$ bar at ~4.1 Ga (Cassata et al. 2012, Kurokawa et al. 2018).

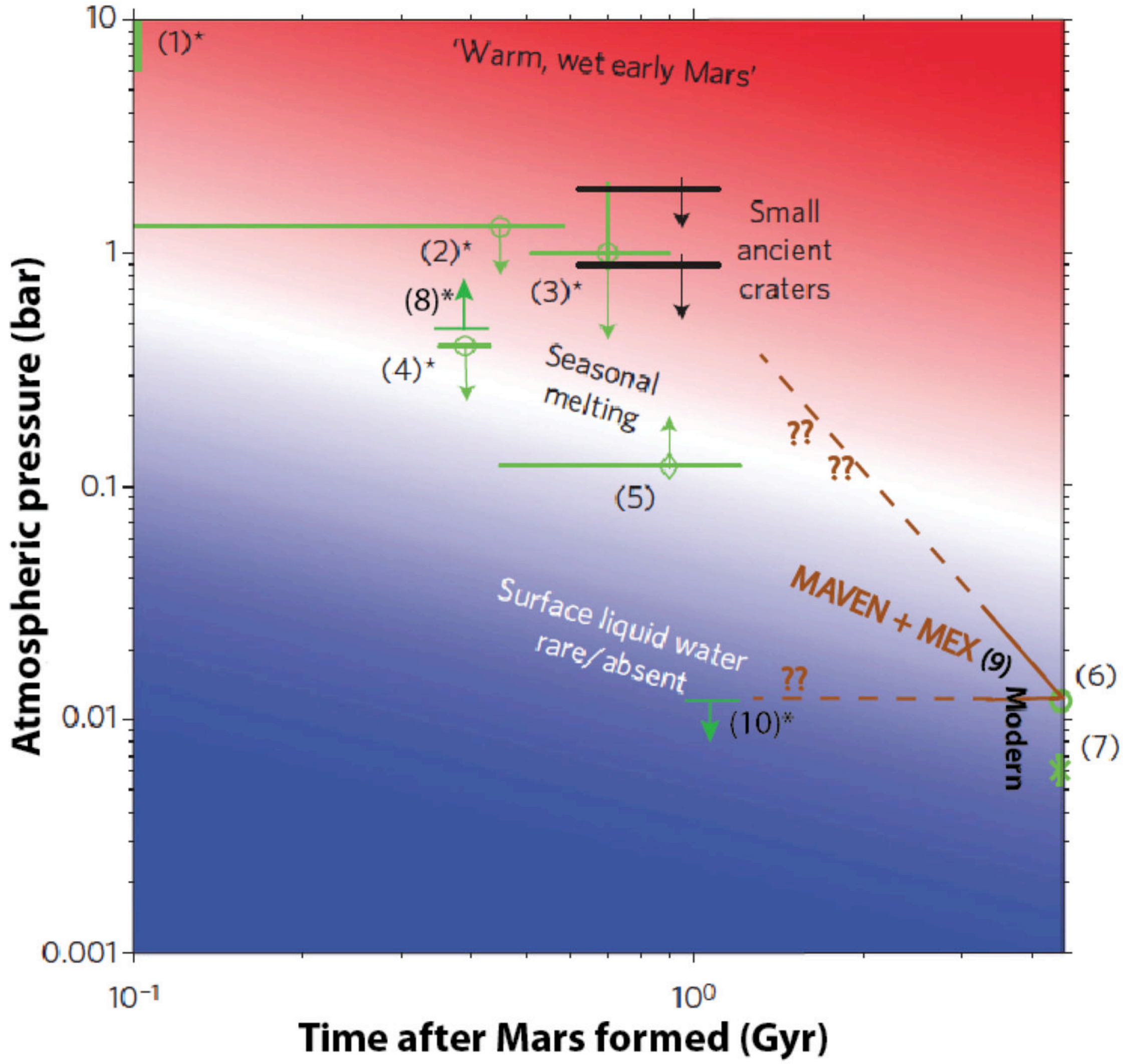

**Fig. 9.** Paleo-atmospheric pressure figure, updated from Kite et al. (2014). Black symbols show result of Kite et al. (2014). * = indirect estimate. (1*) estimate of initial atmospheric $pCO_2$ based on cosmochemistry; (2*) prehnite stability (Ehlmann et al. 2011b, Kite et al. 2014); (3*) carbonate Mg/Ca/Fe (van Berk et al. 2012); (4*) $^{40}Ar/^{36}Ar$ (Cassata et al. 2012); (5) bomb sag (Manga et al. 2012; single data point); (6) modern atmosphere; (7) modern atmosphere + buried $CO_2$ ice; (8*) meteorite trapped-atmosphere isotopic ratios (Kurokawa et al. 2018); (9 – range spanned by brown dashed lines) modern escape-to-space measured by Mars Atmosphere and Volatile Evolution Mission (*MAVEN*) and Mars Express (*MEX*) spacecraft, extrapolated into past (Barabash et al. 2017, Lillis et al. 2017, Brain et al. 2017, Ramstad et al. 2018, Jakosky et al. 2018); (10*) bedform wavelengths (Lapôtre et al. 2016). Approximate and model-dependent implications for climate are shown by background colours.

According to isotopic proxy data from Mars samples, modeling, and analogy to the Lunar record, processes that would tend to reduce Mars' atmospheric pressure were most vigorous before ~(3.9-4.0) Ga (Lammer et al. 2013, Niles et al. 2013). Most of Mars' initially-at-the-surface volatile content escaped to space (Catling 2009, Jakosky & Phillips 2001, Catling & Kasting 2017, Jakosky et al. 2017, Shaheen et al. 2015). However, the most vigorous escape is thought to predate the Hellas impact (Cassata 2017, Catling & Kasting 2017, Webster et al. 2013). It is not clear how much $CO_2$ escape-to-space occurred after the Hellas impact. However, the Hellas impact predates all of the evidence for river-forming climates on Mars and therefore the Hellas impact predates Mars' wet-to-dry transition. Therefore, it is also not clear whether or not $CO_2$ escape-to-space was the major driver of Mars' wet-to-dry climate transition.

### 4.2. Peak warmth: >-4°C (very high confidence), >14°C (medium confidence). Peak mean-annual warm-climate temperature: -18°C to 40°C (low/medium confidence).

Early Mars rivers and aqueous minerals required liquid water to form. Pure liquid water implies temperatures >0 °C. Temperatures $\gtrsim$ -4 °C are needed even taking into account salinity; for example, the most-commonly-detected sulfates on Mars are Mg-sulfates, and the $MgSO_4$-$H_2O$ eutectic is -4 °C. $\gtrsim$(-4 °C) peak temperatures are consistent with mean annual temperatures ($T_{av}$) -18 °C (Doran et al. 2002). At $T_{av}$ = -18 °C in the McMurdo Dry Valleys, Antarctica, lakes are permanently covered with an ice layer. Ice thickness is only <10m (McKay et al. 1985). The thin ice cover is sustained despite low $T_{av}$, because freezing of melt-water supplied during a brief melt season warms the deep parts of the lakes (McKay et al. 1985). Beneath long-lived, perennially-ice-covered lakes, unfrozen ground develops and can allow deep groundwater to exchange with near-surface lake waters even in a cold climate (e.g. Mikucki et al. 2015). Subglacial water flow near the South Pole of Mars ~3.6 Ga requires polar surface temperatures >(-73°C), warmer than today's polar temperature but consistent with low-latitude $T_{av}$ < 0 °C (Fastook et al. 2012). An upper limit on post-4.0 Ga long-term mean temperatures of ($22^{+8}_{-\infty}$) °C has been obtained from Ar diffusion kinetics in ALH 84001 (Cassata et al. 2010). Remanant magnetization in ALH 84001 implies that that meteorite never got hotter than 313K (Weiss et al. 2000). These limits all permit peak mean annual temperatures ($T_{av}$) = 255-313K, which is a wide range.

Within this wide range, there are few firm constraints on Early Mars paleo-temperature. Meteorite evidence indicates >298K at ~3.9 Ga from ALH 84001 isotopologue data (Halevy et al. 2011). This data point could mark a warm early climate, or alternatively an impact-induced heating event (Niles et al. 2009). Consistent with colder temperatures at later times, possible pseudomorphs after meridianiite ($MgSO_4 \bullet 11H_2O$) have been observed in ~3.6 Ga rocks by *Opportunity* (Peterson et al. 2007). Because meridianiite turns into slurry at 275K, this is evidence for annual average temperatures below 275K. *Curiosity* data shows multiple episodes of burial-diagenetic groundwater alteration at Gale (e.g. Nachon et al. 2017, Yen et al. 2017, Frydenvang et al. 2017). Burial-diagenetic groundwater alteration does not preclude near-surface permafrost, because temperatures increase with depth below the surface.

The lack of evidence for icy conditions along the *Curiosity* rover traverse hints at ice-free Hesperian lakes (Grotzinger et al. 2015). These *Curiosity* observations are not decisive however, because it is difficult to distinguish the deposits of perennially-ice-covered lakes from the deposits of ice-free lakes in the sedimentary rock record (Head & Marchant 2014, Rivera-Hernandez et al. in press).

An important piece of evidence in favor of $T_{av}$ > 0 °C on Early Mars is a globally distributed Noachian-aged weathering profile revealed by orbiter data (Carter et al. 2015). This profile is typically several meters thick, and 200 m thick at Mawrth (Loizeau et al. 2012, Carter et al. 2015, Loizeau et al. 2018). The profile includes an Al-phyllosilicate, likely formed by leaching of basaltic-composition materials. The Al-phyllosilicate overlies smectite clay. This weathering profile suggests a warm climate. The simplest interpretation of the smectite clay profiles (e.g. Mawrth) is ~$10^6$ yr of warm climate, with summer maxima of (30-40)°C (Bishop & Rampe 2016, Bishop et al. 2018). This stratigraphic interval (or intervals) might correspond to the warmest/wettest climate in Mars' stratigraphically-legible history (Bibring et al. 2006, Carter et al. 2015, Le Deit et al. 2012). No rover has systematically explored the Noachian weathering profiles. Mawrth could be a window into a climate that might predate the valley networks, and might have been more habitable (or habitable for longer) even than the valley-network-forming climate (Bishop et al. 2018, Bishop et al. 2013; §6.2). However, this interpretation might be wrong: acid alteration can form Al-phyllosilicates from basalt at 273K (Zolotov & Mironenko 2007, Zolotov & Mironenko 2016, Peretyazkho et al. 2018, Loizeau et al. 2018). Moreover, clays exist in the Coastal Thaw Zone of the McMurdo Dry Valleys (Kaufman et al. 2018).

To sum up, we know little about Early Mars paleotemperature. $T_{av}$ > 0 °C is supported by the weathering profiles that feature Al-phyllosilicates, the isotopologue data from ALH 84001, and the ubiquitous evidence for groundwater flow. However, when the data are considered together, the alternative hypothesis of $T_{av}$ ≤ -13 °C is no less likely.

## 5. Data-model comparison.

Table 1 sums up the geologic constraints, and Table 2 compares constraints to models.

## 5.1. Challenges and opportunities for models (Table 2).

Explaining rivers and lakes on Early Mars is difficult. Only in the past few years have models emerged that can self-consistently account for rivers and lakes on Early Mars. However, also in the last few years, new geologic constraints have been published (Table 1). No single published model can match all the geologic constraints without special circumstances (Table 2). For example,

- The lower bound on post-3.4 Ga runoff, combined with the fact that almost all large impacts on Mars predate 3.4 Ga (Irwin et al. 2013), means that direct conversion of impact kinetic energy into latent heat of water melting/vaporization gives too little liquid water to explain the data (Steakley et al. 2017, Turbet et al. 2017a, Segura et al. 2013).

- Single, isolated super-volcanic eruptions would last < 30 yr (Halevy & Head 2014), giving warming pulses too short to match the lake-lifetime constraint.
- Most models assume ≳1 bar of $CO_2$ to provide a baseline of greenhouse warming (e.g. Wordsworth et al. 2017). It is not clear whether or not such a thick atmosphere existed by the time Gale crater sediments formed. If atmospheric pressure was ≲ 0.1 bar by the time Gale crater sediments formed, then of currently proposed global models, most would not explain the Gale data. Melting snow and ice should be especially difficult for $P$ < 0.1 bar (Hecht 2002). Therefore, more constraints on paleopressure would be valuable.
- Wet climates on Mars existed ≲3.4 Ga. Post-3.4 Ga persistence is a challenge for models, because a thin atmosphere is expected and a thin atmosphere makes melting difficult (e.g. Wordsworth et al. 2017).

Therefore, explaining rivers and lakes on Early Mars remains difficult. This difficulty, however, also creates opportunities for models. For example,

- Surface temperatures >0 °C are not enough! To be compared to the "wet-pass-filtered" geologic record (§3.1), models should account for evaporitic cooling, evaporative removal of snowpack, and meltwater infiltration and refreezing (e.g. Clow 1987, Woo 2012, Dingman 2014, Williams et al. 2009).
- Mars could have had a climate with global average annual average $T_{av}$ >0 °C. However, the persistence of easily-dissolved minerals rules out a globally warm, wet climate having occurred for more than a few percent of Mars' post-Noachian history. The data are consistent with no such climate ever having occurred except for a couple of years after very large impacts. Indeed, the data can be matched with $T_{av}$ ≲ -13 °C (Fairén 2010). This relaxed target is easier to match in models.
- The cumulative duration of river-forming climates could be <$10^7$ yr. Indeed a ≫$10^8$-yr long Earth-like climate would overpredict both weathering and erosion. This opens the door to warming mechanisms that rely on infrequent, but expected, triggers.
- Different warming mechanisms may have been active at different times. For example, $H_2$ outgassing was likely more potent in the Early/Mid Noachian than in the Late Hesperian / Amazonian. The { $T_{av}$ - $P$ – $\tau$ – $R$ – $I$ – $N$ } target for the Noachian/ Hesperian boundary is not the same as the { $T_{av}$ - $P$ – $\tau$ – $R$ – $I$ – $N$ } target for the Late Hesperian / Early Amazonian.
- The $H_2O$-ice cloud greenhouse model is not rejected by any geologic data (Table 2).

## 5.2. "False friends": ambiguous proxies

Some geologic data are valid on their own merits, but easy to over-interpret translating between data and models of Early Mars climate. These "false friends" include:

(1) Large gaps in the latitude-longitude distribution of valley networks in THEMIS and MOLA databases may be preservation artifacts. Most large gaps in the latitude-longitude distribution of valley networks in THEMIS and MOLA databases (Hynek et al. 2010, Luo et al. 2017) correspond to post-fluvial resurfacing (i.e., Amazonian lava or Amazonian ice-rich mantling deposits). The biggest exception, Arabia Terra, has numerous river deposits at scales too fine to be detected in THEMIS or MOLA (Hynek et al. 2010, Davis et al. 2016, Williams et al. 2017). Given that large gaps in the valley network distribution are mostly preservation artifacts, conclusions that rest on large gaps in the valley network distribution are questionable. However, a climate model that predicts zero precipitation in a zone where data show that precipitation-fed rivers did occur can be ruled out (as done by Wordsworth et al. 2015). Moreover, other aspects of the spatial distribution of valleys are potentially useful constraints. For example, valleys are deepest near the equator (Williams et al. 2001). Moreover, clumps in the distribution of large alluvial fans (Fig. 1, Kraal et al. Icarus 2008a) are not a preservation artifact. These clumps remain completely unexplained.

(2) <u>There is no unambiguous geologic evidence for post-Hellas True Polar Wander (TPW).</u> For example, ice-cap offset (Kite et al. 2009) can be explained by atmospheric dynamics instead of TPW (Scanlon et al. 2018). Deflection of candidate shorelines (Perron et al. 2007) can be explained by flexure instead of TPW (Citron et al. 2018).

(3) <u>Neither high drainage density, nor softened crater rims, need imply rainfall.</u> High drainage density (>5 $km^{-1}$ of total channel length per unit area) is sometimes observed on Mars. High drainage density has been claimed to be “a gun, although not a smoking gun” for past rainfall (Malin et al. 2010). However, high drainage density can be seen in some snowmelt landscapes (Kite et al. 2011b). Therefore, high drainage density does not constrain the phase of precipitation (rain versus snow/ice melt). Some workers assert that the softened appearance of ancient crater rims implies rainfall (Craddock & Lorenz 2017, Ramirez & Craddock 2018). This is not correct: rainsplash may soften hillslopes under some circumstances (Dunne et al. 2016), but rounded hill-crests can form due to non-rainfall processes (Melosh 2009). Therefore, the phase of precipitation (rain vs snow/ice melt) cannot be “read” from rounded hill-crests alone. Rainfall can be proven by fossilized rainsplashes – but these have not yet been observed at Mars.

(4) <u>Km-deep precipitation-fed canyons can form in <$10^6$ yr.</u> Many precipitation-fed Mars river valleys are cut >$10^2$ m deep into bedrock; some are $10^3$ m deep. The only $10^3$ m deep and >$10^2$ km long bedrock canyon exposed on Earth – the Grand Canyon of the Colorado – took 5-70 Myr to carve – implying a rate of $10^{-4}$ m/yr (Flowers & Farley 2012, Darling & Whipple 2015). However, it is incorrect to infer that Mars’ bedrock river canyons must also have taken $10^7$ yr to form. To the contrary, much faster fluvial erosion of bedrock occurs both in the laboratory and in the field (Whipple et al. 2000, Dethier 2001, Anderson & Anderson 2010, Hildreth & Fierstein 2012, Gallen et al. 2015). This is especially true for basin breach flooding events (Baker & Kale 1998). To figure minimum Mars canyon erosion timescales, it is better to drop the Earth analogy approach, and instead use timescales for sediment-transport-limited erosion (Lamb et al. 2015). This is a better approach because sediment transport is better understood than bedrock erosion. Using the sediment-transport-limited estimation procedure, Mars valley formation timescales can be as short as $10^4$ yr (Hoke et al. 2011, Rosenberg et al. 2019).

(5) More $H_2O$ need not imply a wetter climate. Mars climate need not be sensitive to gradually-imposed factor-of-a-few increases in the amount of $H_2O$ substance on Mars' surface. The modern Mars atmosphere+(ice cap) system has ~34 m Global Equivalent Layer of $H_2O$ ice (Carr & Head 2015), $10^{-5}$ m Global Equivalent Layer of $H_2O$ vapor (Smith 2002), and no liquid water. D/H of 3.3±0.3 Ga rocks shows that H corresponding to ~60m Global Equivalent Layer of $H_2O$ has been lost to space since 3.3±0.3 Ga (Mahaffy et al. 2015) [5]. The implication, consistent with glacial geologic data (Scanlon et al. 2018), is that Mars in the Hesperian had $H_2O$-ice sheets that were thicker. Counterintuitively, this does not imply a wetter climate. The tiny fraction of $H_2O$ in Mars' atmosphere is regulated by lag-deposit formation and by ice-cap albedo (Mischna & Richardson 2005). Ice sheet size is only a secondary factor. Volatile abundance does becomes important for climate when ice sheets spread under their own weight to cover the planet (Turbet et al. 2016), but even >700 m Global Equivalent Layer of water is not enough to put Mars into that regime (Fastook & Head 2015). To sum up, tripling the amount of $H_2O$ substance on the surface of Mars – with no other changes – would just lead to thicker ice caps, and not a liquid ocean.

For many Early Mars data, the data allow for multiple interpretations: one that is familiar from Earth experience, and alternatives that (while consistent with basic theory and with experiments) would require processes to operate differently on Mars than on Earth. These alternative, strange explanations can turn out to be true. For example, High Resolution Imaging Science Experiment (HiRISE; McEwen et al. 2007) monitoring of active gully-shaping processes shows that the currently active agent is $CO_2$ (Diniega et al. 2013, Dundas et al. 2017a). $CO_2$-driven gully modification would not have been anticipated from Earth analogy. Gully-shaping processes on modern Mars are a case where Earth analogy led to premature conclusions. This is one reason why the method of Earth analogy is a questionable method for testing the hypothesis that an Earth-like climate existed on Early Mars.

## 5.3. Key parameters for Early Mars climate: unifying frameworks in time and space

Unifying frameworks in space. For a given time slice, climate-driven geologic activity for Early Mars can be plotted using latitude/elevation coordinates (Fig. 5, Appendix Fig. A.1). In Fig. 5, we plot precipitation-fed rivers and alluvial fans. We also plot light-toned layered sedimentary rocks, which are generally believed to record warmer/wetter past climates (Malin & Edgett 2000, Grotzinger et al. 2005, Kite et al. 2013, Andrews-Hanna & Lewis 2011; but see Niles et al. 2017). The data indicate latitude-elevation preferences, and suggest trends over time (§2.3). These observations can be related to past climate, as follows. Latitude and elevation are key parameters for models of Early Mars climate-driven geologic activity. Snow or ice will melt if they get warmed above 0 °C. Warming can be supplied by (1) insolation and (2) by turbulent exchange with the air. Insolation (for a prescribed slope and aspect) is mainly controlled by latitude. Insolation is the main warming agent on Mars today, because the air is so thin. But for atmospheric pressure $\gtrsim$ 0.3 bar, turbulent exchange with the air is the main control on surface temperature (Wordsworth 2016). That is because the turbulent fluxes that exchange heat between the surface and adjacent air are proportional to air density. Air temperature decreases with elevation, so at high pressure, turbulent fluxes ensure that surface temperature also decreases with elevation. Moreover, at high pressure, greenhouse warming becomes relatively more important (for all elevations), and greenhouse warming is strongest at low elevation. Thus, at high pressure, elevation and latitude are both important for determining surface temperature. This is consistent with data for the Noachian (e.g. Fig. 5b).

[5] This suggests that most of the 3-Gyr-integrated O loss inferred from *MAVEN* (Lillis et al. 2017) was "paired" with H, and therefore that $CO_2$ escape from Mars over the last 3.5 Gyr was $\ll$0.8 bar.

<u>Unifying frameworks in time.</u> Climate-driven geologic activity for Early Mars can be plotted as a wetness probability distribution function. Suppose we are transported back in time to a random year in Early Mars history. How wet would be expect it to be during that year? How often would the energy available for melting of snow and ice exceed (0-15) W/m$^2$ - allowing soil-wetting and olivine dissolution? How often would the energy available for melting of snow and ice be high enough for surface runoff and fluvial sediment transport? (We arbitrarily pick 50 W/m$^2$ energy available for melting as the threshold for runoff; see Clow 1987 for more detailed calculations). We plot our expectations, based on the constraints from Table 1, on a "wetness probability distribution function" (Fig. 10).

From the wetness pdf, we deduce the following:

(1) No data rule out the hypothesis that Early Mars was globally dry for >90% of years.

(2) The apparent discrepancy between olivine persistence and the time needed for sedimentary-rock build-up (Fig. 10) can be understood as follows. To pile up a sedimentary rock mound, less-weathered minerals are blown in by the wind. This import of less-weathered materials resets the olivine-dissolution stopwatch. Any one sediment layer within the mound "sees water" for $\ll 10^7$ yr. Therefore, each sediment layer can retain olivine. Despite this, the sedimentary rock column as a whole records the availability of water $>10^7$ yr.

(3) Combining the post-Noachian olivine persistence constraint with the post-Noachian alluvial fan formation constraint indicates that it was not wet very often, but when it was wet, it was often very wet. Surprisingly, this inference from data can be matched by a basic surface energy-balance model of long-term climate evolution that includes only $CO_2$, solar luminosity, and chaotic obliquity forcing (Mansfield et al. 2018).

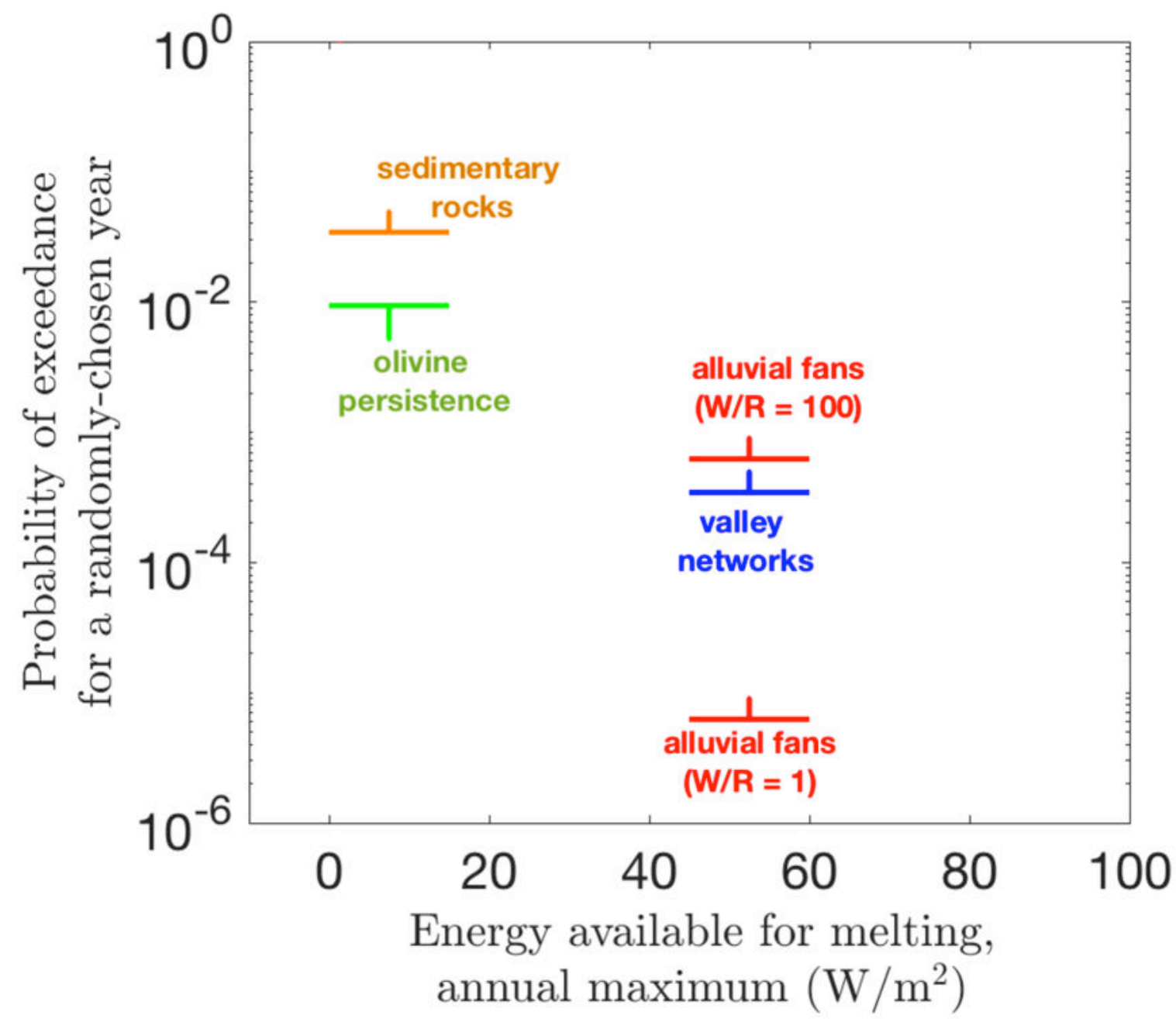

**Fig. 10.** The wetness probability distribution function (wetness pdf) for Early Mars. The x-axis corresponds to energy available for snow/ice melt (i.e., net surface energy balance for a snow/ice surface at 273.15K). 0.5 m/yr runoff in each year with some runoff is assumed for the build-up of the alluvial fans.

| | Confidence level (VH = very high, H = high, M = medium, L = low) | $CO_2$ + $H_2O$(v) greenhouse + orbital forcing (*a*) | Lower surface albedo (*b*) | MEGAOUTFLO (*c*) | Direct conversion of impact energy to $H_2O$ vapor (*d*) | Localized Palimpsest (*e*) | Volcanic eruptions / $SO_2$ (*f*) | $H_2O$-ice cloud greenhouse (*g*) | Bursty release of $CH_4$ or $H_2$ from subsurface (*h*) | Impact delivery of $CH_4$ or $H_2$ (*i*) | Steady release of $H_2$ (±$CH_4$) from subsurface (with limit cycling) (*j*) | Steady release of $H_2$ (±$CH_4$) from subsurface (no limit cycling) (*k*) |
|---|---|---|---|---|---|---|---|---|---|---|---|---|
| Post 3.4 Ga - precipitation-sourced water column >1 km | H (§2.1) | ✗ | ✗ | ? | ✗ | ✗ | ? | ✓ | ? | ✗ | ? | ? |
| Time spanned by river-forming climates ≫$10^8$ yr | H (§2.2) | ✗ | ✗ | ✓ | ? | ✓ | ✓ | ✓ | ? | ? | ✓ | ✓ |
| Number of river-forming periods ≥ 2 | H (§2.2) | ✗ | ✗ | ✓ | ✓ | ✓ | ✓ | ✓ | ? | ✓ | ✓ | ✗ |
| Ocean size at maximum extent >1.1 × $10^6$ km$^2$ | VH (§2.9) | ✗ | ✗ | ✓ | ? | ✗ | ✗ | ? | ✓ | ✓ | ✓ | ✓ |
| Last year of flow in the rivers of Mars <3.4 Ga | H (§2.2) | ✗ | ✗ | ✓ | ✗ | ✓ | ✓ | ✓ | ? | ? | ✗ | ✗ |
| Duration (τ) of longest individual river-forming climate > ($10^2$ - $10^3$) yr | M (§2.5) | ✗ | ✗ | ? | ✗ | ✗ | ✗ | ✓ | ✓ | ✓ | ✓ | ✓ |
| Peak runoff production ( *R*) >0.1 mm/hr | H (§2.6) | ✗ | ✗ | ✓ | ✓ | ✓ | ? | ? | ? | ✓ | ✓ | ✓ |
| Intermittency during wet events (*I*): peak runoff production <10% of the time | H (§2.7) | ? | ? | ? | ? | ✗ | ✓ | ✓ | ✓ | ✓ | ✓ | ✓ |
| Cumulative wet years during valley-network-forming climate episode ($\tau \times N$) >$10^5$ yr | M (§2.8) | ✗ | ✗ | ? | ✗ | ✗ | ? | ✓ | ? | ? | ✓ | ✓ |
| Duration of surface liquid water for "average" post-Noachian Mars <$10^7$ yr | M/H (§3.3) | ✓ | ✓ | ✓ | ✓ | ✓ | ✓ | ✓ | ✓ | ✓ | ? | ✗ |
| Years of sediment deposition in sedimentary rock record >$10^7$ yr | H (§3.1) | ? | ? | ✗ | ✗ | ✗ | ✓* | ? | ✗ | ✗ | ✓ | ✓ |
| Time span of deposition for layered, indurated, equatorial sediments ≫$10^8$ yr | H (§3.1) | ? | ? | ✓ | ? | ? | ✓ | ✓ | ? | ✗ | ✓ | ✓ |
| Water column required to indurate sedimentary rocks > 20 km | M/H (§3.2) | ? | ? | ? | ✗ | ✗ | ✓ | ✓ | ? | ? | ? | ? |
| Paleo-atmospheric pressure <(1-2) bar | L/M (§4.1) | ? | ✓ | ✓ | ? | ✓ | ? | ✓ | ? | ? | ? | ? |
| Peak annual-mean warm-climate temperature at river locations ($T_{av}$) > -18°C | L (§4.2) | ? | ? | ✓ | ✓ | ? | ? | ? | ? | ? | ✓ | ✓ |

Table notes:- *a*: Pollack et al. 1987, Forget et al. 2013, Kite et al. 2013, Mansfield et al. 2018. *b:* Fairén et al. 2012. *c*: Baker et al. 1991. Assuming massive $CH_4$ release during outflows: $H_2O$ alone is not sufficient – see Turbet et al. 2017b. *d:* Segura et al. 2008, Toon et al. 2010. *e:* Gulick & Baker 1989, Kite et al. 2011a, 2011b. *f:* Tian et al. 2010, Halevy & Head 2014, Kerber et al. 2015. *g*: e.g. Urata & Toon 2013, Ramirez & Kasting 2017, Kite et al. 2018b. Possibly impact-triggered. *h:* Kite et al. 2017a, Tosca et al. 2018. *i*: Haberle et al. 2018 *j*: Batalha et al. 2016. *k*: Ramirez 2017. *: See Niles et al. 2017.

**Table 2.** Comparison of Early Mars geologic constraints to selected models. ✓ = this model can explain this observation; ? = special, but plausible circumstances may be required to produce this observation from this model; ✗ = this model does not plausibly explain this observation.

## 6. Summary.

## 6.1. A speculative hypothesis for Early Mars climate.

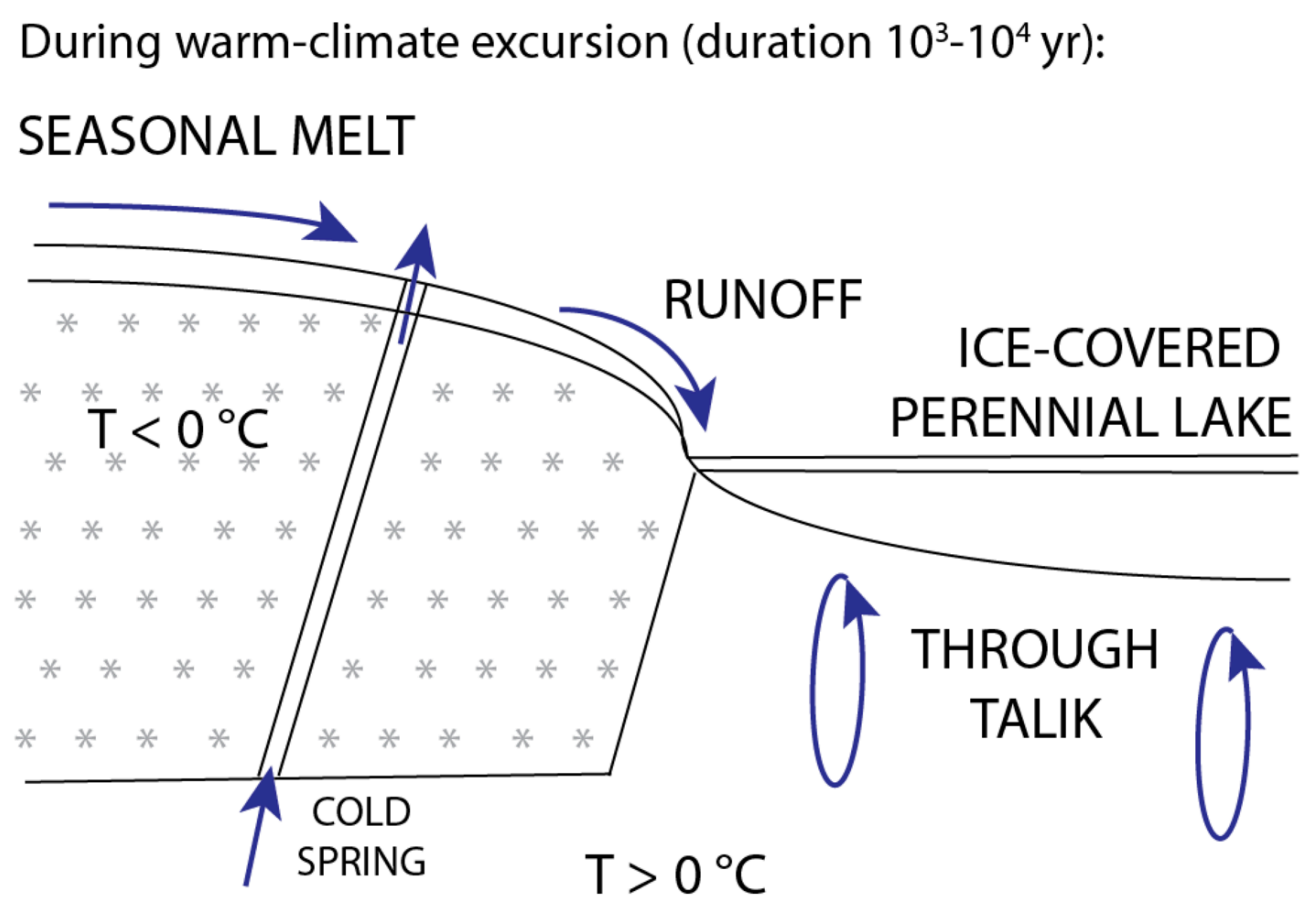

**Fig. 11.** A speculative hypothesis for Early Mars climate (see, e.g. McKay et al. 2005, Head et al. 2017).

The data reviewed above are mostly consistent with the following speculative hypothesis for Early Mars climate. This hypothesis follows the ideas of McKay et al. (2005), and is similar to the Late Noachian Icy Highlands hypothesis (Head et al. 2017). This hypothesis is not a consensus statement, it is controversial, and it is intended to spur further work.

*"During the middle Noachian through early Amazonian, Mars experienced individually prolonged, but increasingly infrequent excursions to temperatures as warm as places near McMurdo, Antarctica today – perhaps as warm as Central Siberia. During these excursions, perennial lakes, filled by seasonal meltwater runoff, existed beneath thin (<10 m) ice cover. Taliks beneath these lakes, as well as conduits through permafrost that were maintained either by high solute concentration or by advection, permitted surface-interior hydrologic circulation (Schiedegger & Bense 2014). Warmer-than-Central-Siberia temperatures occurred only in the immediate aftermath (<$10^2$ yr) of basin-forming impacts – these impact-generated transients were too brief to permit interior-to-surface groundwater flow."*

The above hypothesis matches much, but not all of the data (e.g. Bishop et al. 2018). This hypothesis is acceptable to many palates: many climate models can achieve McMurdo Dry Valleys-like conditions (e.g. Wordsworth et al. 2017, Kite et al. 2017). The above hypothesis can also reproduce the best-understood geologic data. Because of the key role of sub-lake 'through taliks,' the above hypothesis also permits both vertical segregation and vertical integration of the Early Mars hydrosphere (e.g., Head 2012). On the other hand, many climate models predict climates that were intermittently (or stably) warmer than the McMurdo Dry Valleys (Urata & Toon 2013, Batalha et al. 2016, Ramirez 2017); conversely, some climate models predict that lake-enabling conditions were <$10^2$ yr duration (Segura et al. 2013). Moreover, *Curiosity* data show no evidence in favor of the proposed subzero conditions (Rivera-Hernandez et al. in press). Therefore, more tests are needed. These future tests might include runoff production, and a search for evidence of ancient low-latitude snow/ice melt. A particularly useful measurement would be grain-size data for ancient fluvial sediments. Measuring the size of clasts moved by rivers has been a key justification for pushing orbiter imagery to higher resolution in the past (Malin et al. 2010, McEwen et al. 2007). Today, we are frustratingly a factor of a few below orbital detection of the most relevant grain sizes – gravel (Dietrich et al. 2017). Future methods might include active methods for surface roughness characterization (Pitman et al. 2004).

## 6.2. Implications for the Mars Exploration Program.

Most people on Earth have put money towards Mars missions. Mars missions have been initiated by China, India, Europe, Russia, Japan, the United Arab Emirates, and the United States. The United States has supplied the largest investment so far, and all eight of the successful landed missions. NASA objectives for Mars exploration are defined in the Mars Exploration Program Analysis Group Science Goals Document (MEPAG 2018).

Goal II of the Mars Exploration Program is "Understand the processes and history of climate on Mars." The data reviewed in this paper show great promise for progress on this goal. Mars' geology records distinct, separable climate regimes spanning billions of years. Therefore, Mars records a rich set of natural experiments for understanding how planets in general behave (Ehlmann et al. 2016). Models of long-term planetary climate evolution can be tested through continued data collection and analysis.

However, Goal I of the Mars Exploration Program is "Determine if Mars ever supported life." The data reviewed in this paper highlights a major concern for this goal. Specifically:

- Mars's surface could have supported life during multiple time windows during its history (e.g. Grotzinger et al. 2014, Knoll et al. 2005, Squyres et al. 2008), sprinkled across >1 Gyr. However, the data do not require >1 Gyr of continuous Mars surface habitability. To the contrary, a minimal model in which globally surface-sterilizing conditions occurred >90% of the time, even on Early Mars, can match data. Gappy and spotty surface habitability would be a challenge for the persistence of surface life (Cockell 2014, Westall et al. 2015)[6].

- Maximizing the probability of finding life beyond Earth, as well as the science value of a negative result, requires investigating rocks that date from Mars' most-habitable period. Intermittent surface habitability after ~4.0 Ga may mean that the most habitable times in Mars surface history predate the interpretable-from-orbit geologic record. Most of Mars' crust (and thus most of Mars' volcanic activity) predates the Hellas impact (Taylor & McLennan 2009), and Mars' geodynamo shut down before the Hellas impact (Lillis et al. 2013). Theoretically, Mars should have had abundant liquid water right after the planet formed (e.g., Cannon et al. 2017). However, landing sites for rovers and sample return missions are chosen based on orbiter data. Use of orbiter data biases rovers to land at sites that are easier to interpret from orbit. Sites that are easier to interpret from orbit tend to be relatively young. But the most habitable times on Mars history may simply not leave a record that can be read from orbit today, because the earliest record has been jumbled by impacts and volcanism (Cannon et al. 2017, Andrews-Hanna & Bottke 2017). Because the data are consistent with a scenario in which the post-Hellas surface was not continuously habitable, the understandable bias towards post-Hellas geologic targets for landed missions is a serious concern. This concern may be mitigated by including megabreccia samples in a return cache (McEwen et al. 2009, Weiss et al. 2018).

[6] Research aimed at Goal I is focused on surface life. That is because the search for ancient microbial fossils of Earth's surface biosphere has a >50 year record of developing techniques that may be applied to Mars sediments (McMahon et al. 2018), whereas techniques for finding fossils of deep subsurface life are less well-developed (Onstott et al. 2018).

**Acknowledgements.** The results listed above sum up the work of thousands of engineers and scientists. Many great papers are omitted from this review for concision. I am grateful to Chris McKay and Caleb Fassett for formal reviews, and to Tim Goudge, Paul Niles, and Brian Hynek for informal read-throughs. I thank David P. Mayer for generating the CTX DTM used in Fig. 2, and Jack Mustard for sharing a preprint. This paper was stimulated by the Fourth International Conference on Early Mars, and I thank the organizers and participants for that meeting. This work was funded in part by the U.S. taxpayer, via NASA grant NNX16AJ38G.

## Appendix.

| Epoch | Ages of epoch |
|---|---:|
| Late Amazonian | 0.27-0 Ga |
| Middle Amazonian | 1.03-0.27 Ga |
| Early Amazonian | 3.24-1.03 Ga |
| Late Hesperian | 3.39-3.24 Ga |
| Early Hesperian | 3.56-3.39 Ga |
| Late Noachian | 3.85-3.56 Ga |
| Middle Noachian | 3.96-3.85 Ga |
| Early Noachian | ~4.0-3.96 Ga |
| pre-Noachian | >4.0 Ga |

**Table A1.** Absolute date estimates used in this paper, reproduced from Table 1 of Michael (2013) which in turn follows the Hartmann (2005) chronology.

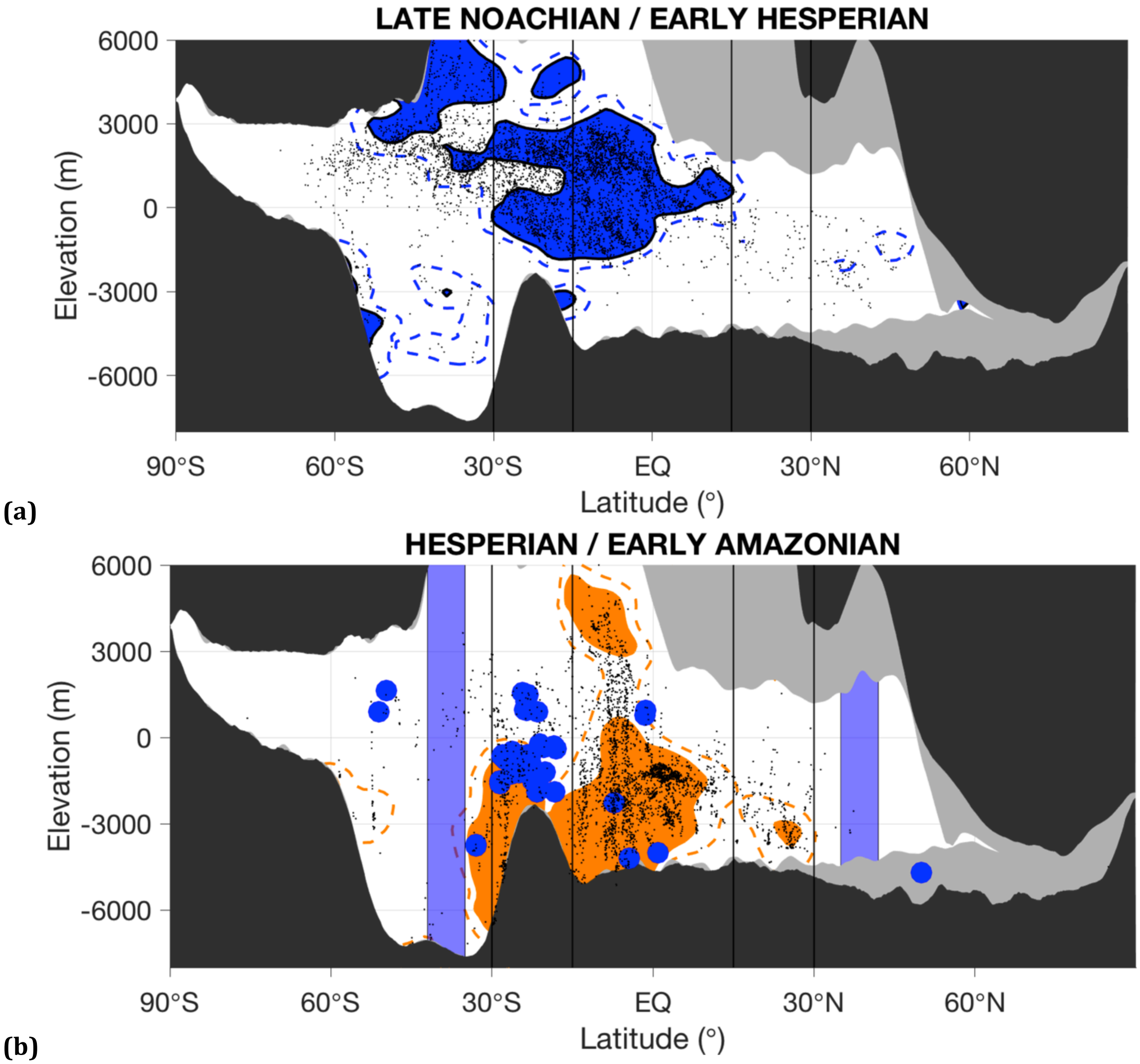

**Fig. A1.** Latitude-elevation plots of climate-relevant geologic activity for the Late Noachian / Early Hesperian and the Hesperian / Early Amazonian. The black regions have no data, and the gray regions correspond to terrain that was geologically reset after the time slice in question. Vertical black lines highlight latitudes ±15° and ±30°. **(a) Late Noachian / Early Hesperian time slice:** Black dots correspond to individual valleys from the catalog of Hynek et al. (2010). Only every 10th valley is plotted, for legibility. The density of black dots reflects the nonuniform distribution of elevation as a function of latitude (for example, there is not much Noachian terrain S of 30°S above +3 km elevation). To correct for this effect, and get the latitude-and-elevation dependent density of valleys, we used a kernel density estimator. The resulting blue zone corresponds to the latitude/elevation zones that have the highest density of valleys, and is drawn to contain ⅔ of the valleys. Blue dashed line is the same, but for 9/10 of the valleys. **(b) Late Hesperian / Amazonian time slice:** Blue disks mark large alluvial fans (combining catalogs of Moore & Howard 2005 and Kraal et al. 2008a). Pale blue stripes mark latitude range of Fresh Shallow Valleys (Wilson et al. 2016). Black dots

correspond to the sedimentary rocks from the catalog of Malin et al. (2010). The density of black dots reflects the nonuniform distribution of elevation as a function of latitude. To correct for this effect, and get the latitude-and-elevation dependent density of sedimentary rocks from Malin et al.'s (2010) catalog, we used a kernel density estimator. The resulting orange zone corresponds to the latitude/elevation zones that have the highest density of sedimentary rocks, and is drawn to contain ⅔ of the sedimentary rocks. Orange dashed line is the same, but for 9/10 of the sedimentary rocks.

## References.